\documentclass[twocolumn,showpacs,preprintnumbers,nofootinbib,prd,superscriptaddress,groupedaddress,10pt]{revtex4-1}

\usepackage{graphicx,amssymb,amsmath,amsthm,amsfonts,epsfig,epsf}
\usepackage[linktocpage]{hyperref}
\usepackage[usenames]{color}
\usepackage{epstopdf}

\usepackage{aas_macros}
\usepackage{bm}
\usepackage{dcolumn}
\usepackage[latin1]{inputenc}
\usepackage{latexsym}
\usepackage{rotating}
\usepackage{hyperref}
\usepackage{color}
\usepackage{longtable}
\usepackage{enumerate}
\usepackage{tensor}
\usepackage{url}
\def\nn{\nonumber}

\newcommand{\ben}{\begin{enumerate}}
\newcommand{\een}{\end{enumerate}}

\newcommand{\Lap}{\bigtriangleup}

\newcommand{\ldif}{{\rm d}_t}
\newcommand{\del}{\partial}

\def\be{\begin{equation}}
\def\ee{\end{equation}}
\newcommand{\beq}{\begin{eqnarray}}
\newcommand{\eeq}{\end{eqnarray}} 
\newcommand{\ba}{\begin{align}}
\newcommand{\ea}{\end{align}}

\def\l{\left}
\def\r{\right}

\def\nn{\nonumber}

\DeclareMathOperator\erf{erf}

\newcommand{\dif}{{\rm d}}

\newcommand{\tgam}{\tilde{\gamma}}
\newcommand{\tGam}{\tilde{\Gamma}}

\newcommand{\tA}{\tilde{A}}

\newcommand{\bA}{\bar{A}}

\def\Lie{\mathcal{L}}

\def\H{\mathcal{H}}
\def\M{\mathcal{M}}

\def\ba{\bar{a}}

\def\bA{\bar{A}}

\def\tA{\tilde{A}}

\begin{document}

\title{Interaction between bosonic dark matter and stars}

\author{
Richard Brito$^{1}$,
Vitor Cardoso$^{1,2,3}$, 
Caio F. B. Macedo$^{1}$,
Hirotada Okawa$^{4,5}$,
Carlos Palenzuela$^{6}$
}
\affiliation{${^1}$ CENTRA, Departamento de F\'{\i}sica, Instituto Superior T\'ecnico -- IST, Universidade de Lisboa -- UL,
Avenida Rovisco Pais 1, 1049 Lisboa, Portugal}
\affiliation{${^2}$ Perimeter Institute for Theoretical Physics, 31 Caroline Street North
Waterloo, Ontario N2L 2Y5, Canada}
\affiliation{${^3}$ Dipartimento di Fisica, ``Sapienza'' Universit\`a di Roma \& Sezione INFN Roma1, P.A. Moro 5, 00185, Roma, Italy}
\affiliation{${^4}$ Yukawa Institute for Theoretical Physics, Kyoto University, Kyoto, 606-8502, Japan}
\affiliation{${^5}$ Advanced Research Institute for Science \& Engineering,
Waseda University, 3-4-1 Okubo, Shinjuku, Tokyo 169-8555, Japan}
\affiliation{${^6}$Departament  de  F\'{\i}sica $\&$ IAC3,  Universitat  de  les  Illes  Balears  and  Institut  d'Estudis
Espacials  de  Catalunya,  Palma  de  Mallorca,  Baleares  E-07122,  Spain}

\begin{abstract}
We provide a detailed analysis of how bosonic dark matter ``condensates'' interact with compact stars, extending significantly the results of a recent Letter~\cite{Brito:2015yga}.
We focus on bosonic fields with mass $m_B$, such as axions, axion-like candidates and hidden photons. 
Self-gravitating bosonic fields generically form ``breathing'' configurations, where both the spacetime geometry and the field oscillate, and can interact and cluster at the center of stars.
We construct stellar configurations formed by a perfect fluid and a bosonic condensate, and which may describe the late stages
of dark-matter accretion onto stars, in dark matter-rich environments. These composite stars oscillate at a frequency which is a multiple of $f=2.5\times 10^{14}\,\left(m_{B}c^2/eV\right)\,{\rm Hz}$. 
Using perturbative analysis and Numerical Relativity techniques, we show that these stars are generically stable, and we provide criteria for instability.
Our results also indicate that the growth of the dark matter core is halted close to the Chandrasekhar limit. We thus dispel a myth concerning dark matter accretion by stars: dark mater accretion does not necessarily lead to the destruction of the star, nor to collapse to a black hole. Finally, we argue that stars with long-lived bosonic cores may also develop in other theories with effective mass couplings, such as (massless) scalar-tensor theories.
\end{abstract}



\maketitle

\tableofcontents
\section{Introduction}

There is now compelling evidence that most of the mass and energy enclosed in our Universe is made of yet unseen forms of matter and energy. For the standard picture of the evolution and structure of the Universe to agree with observational data, about a quarter of the Universe must come under the form of nonbaryonic matter, generically called dark matter (DM).
The evidence for DM in observations is overwhelming, starting with galaxy rotation curves,
gravitational lensing and the cosmic microwave background~\cite{Bertone:2004pz}.
While carefully concocted modified theories of gravity can perhaps explain almost all observations,
the most attractive and accepted explanation lies in DM being composed mostly of cold, collisionless particles~\cite{Bertone:2004pz,Klasen:2015uma,Marsh:2015xka}.

Several candidates for dark matter have been proposed~\cite{Bertone:2004pz,Klasen:2015uma,Marsh:2015xka}, of which ultralight bosonic fields, such as axions, axion-like candidates~\cite{Marsh:2015wka,2014NatPh..10..496S,2014PhRvL.113z1302S} or ``hidden photons''~\cite{Ackerman:mha} are an attractive possibility. Axions were originally devised to solve the strong-CP problem, but recently a plethora of other, even lighter fields with masses $10^{-10}-10^{-33}\,{\rm eV}/c^2$, have also become an interesting possibility, in what is commonly known as the axiverse scenario~\cite{Arvanitaki:2009fg}. On the other hand massive vector fields arise in the so-called hidden $U(1)$ sector~\cite{Holdom:1985ag,ArkaniHamed:2008qn,Pospelov:2008jd,Goodsell:2009xc}. This family of candidates are part of what is now referred to as weakly interacting slim particles (WISPs)~\cite{Arias:2012az} in the DM literature.

It is appropriate to emphasize that DM has not been seen nor detected through any of the known standard model interactions.
The only evidence for DM is through its gravitational effect. Not surprisingly, the quest for DM is one of the most active fields of research of this century.
Because DM can only interact feebly with Standard Model particles, and thanks to the equivalence principle, the most promising channel to look for DM imprints consists of gravitational interactions.

Massive bosonic fields minimally coupled to gravity can form structures~\cite{Kaup:1968zz,Ruffini:1969qy,Khlopov:1985jw,Seidel:1991zh,Comment:2015_v2,Guth:2014hsa,Brito:2015pxa}. Self-gravitating {\it complex} scalars may give rise to static, spherically-symmetric geometries called boson stars, while the field itself oscillates~~\cite{Kaup:1968zz,Ruffini:1969qy} (for reviews, see Refs.~\cite{Jetzer:1991jr,Schunck:2003kk,Liebling:2012fv,Macedo:2013jja}). Very recently, analogous solutions for complex massive vector fields where also shown to exist~\cite{Brito:2015pxa}. On the other hand, {\it real} scalars may give rise to long-term stable oscillating geometries, but with a non-trivial time-dependent stress-energy tensor, called oscillatons~\cite{Seidel:1991zh}. Both solutions arise naturally as the end-state of gravitational collapse~\cite{Seidel:1991zh,Garfinkle:2003jf,Okawa:2013jba}, and both structures share similar features.
 
Because the equivalence principle guarantees that all forms of matter gravitate, regions where gravity is strong, such as compact stars, might be a good place to look for signals of DM. Since massive bosonic fields are viable DM candidates~\cite{Bertone:2004pz,Klasen:2015uma}, a natural question is whether they can be accreted inside stars and lead to stable configurations with observable effects. Although not originally framed in the context of DM capture, such solutions, made of both a perfect fluid and a massive complex scalar field, exist~\cite{Henriques:1989ar,Henriques:1989ez,Lopes:1992np,Henriques:2003yr,Sakamoto:1998aj,deSousa:2000eq,deSousa:1995ye,Pisano:1995yk} and can model the effect of bosonic DM accretion by compact stars. Complementary to these studies, accretion of fermionic DM has also been considered, by modeling the DM core with a perfect fluid and constructing a physically motivated equation of state~\cite{Leung:2011zz,Leung:2013pra,Tolos:2015qra}. In a recent \emph{Letter}~\cite{Brito:2015yga} we presented an extension of these results by considering {\rm real} massive scalar and vector fields. Here, we provide some details of the results presented in Ref.~\cite{Brito:2015yga}, while also extending them significantly. 

\subsection{Stars may survive dark matter capture}
Our work may be useful on several fronts. Firstly, our results show that stable DM cores inside compact stars {\it are possible}.
The question of whether these cores are actually formed through dynamical processes in DM environments is harder to answer.
The standard lore -- which our results do not support -- is that the dynamics happen in two stages (see e.g., Refs.~\cite{Goldman:1989nd,Kouvaris:2011fi,Bramante:2014zca,Bramante:2015cua,Kurita:2015vga}):

\noindent {\bf 1.} {\it Accumulation stage,} where DM is captured by the star due to gravitational deflection and a non-vanishing
cross-section for collision with the star material~\cite{Press:1985ug,Gould:1989gw,Goldman:1989nd,Bertone:2007ae}. 
The DM material eventually thermalizes with the star, and accumulates inside a sphere of r.m.s. radius $r_{th}\sim \left(k_BT/\rho_c m_{DM}\right)^{1/2}$, with $k_B$ Boltzmann's constant, $T,\,\rho_c$ the temperature and density of the star, and $m_{DM}$ the mass of the DM particles.

\noindent {\bf 2.} {\it Black hole formation,} after the DM core becomes self-gravitating. The newly formed black hole (BH) eventually eats the host star~\cite{Gould:1989gw,Goldman:1989nd,Kouvaris:2011fi,Bramante:2014zca,Bramante:2015cua,Kurita:2015vga}.

Our results indicate that stable, self-gravitating and self-interacting bosonic DM cores exist and can be explicitly constructed~\cite{Brito:2015yga}, as we detail in the main body of this work. In fact, a similar construction was recently performed for fermionic DM models~\cite{Leung:2013pra}. Thus, the scenario above cannot be generic. In fact, all of these works {\it assume} that gravitational collapse ensues once DM becomes self-gravitating. The stability and evolution of stars is a far more complex affair, and in particular DM dispersion or an increase in the star temperature can easily rule out the collapse scenario. We show that the gravitational cooling mechanism~\cite{Seidel:1993zk} not only disperses bosonic condensates, but it prevents -- generically -- gravitational collapse to occur. 

Our results do not take into account non-gravitational interactions between the star and DM; however, there are strong reasons to suspect that some of the main features are independent, at the qualitative level,
of the nature of the interaction. For example, although our results are formally only valid for zero-temperature bosonic condensates, finite temperature effects are expected to be negligible for bosonic masses much larger than the central temperature of the host star~\cite{Bilic:2000ef,Latifah:2014ima}, such as bosonic fields with masses $\gtrsim$ keV inside old neutron stars or white dwarfs. In fact, Refs.~\cite{Bilic:2000ef,Latifah:2014ima} showed that stable bosonic stars exist for temperatures below a critical temperature which scales linearly with the boson field mass. Finite temperatures tend to increase the radius of the boson star (comparing a star with the same total mass), but do not significantly affect the star's maximum stable mass. In addition, for axionic-type couplings, for instance, all or most of the core energy will be dissipated away under electromagnetic radiation, on relatively small timescales~\cite{Iwazaki:1998eg,Iwazaki:1999my,Iwazaki:2014wka}.

Finally, our study predicts that bosonic DM cores drive the star to 
vibrate at a frequency dictated by the scalar field mass, $f=2.5\times 10^{14}\,\left(m_{B}c^2/eV\right)\,{\rm Hz}$~\cite{Brito:2015yga}, providing a clear means to identify the presence of dark matter in stars, provided these modes are excited to measurable amplitudes.
Helioseismology, developed to the level of a precision science, can now measure individual modes each with an amplitude of $\sim 10\,{\rm cm} \,{s}^{-1}$~\cite{ChristensenDalsgaard:2002ur}. Thus, provided efficient mechanisms exist to gather sufficient DM at the cores of stars, these oscillations will be a smoking gun for DM.

\subsection{Organization of this work}

For the reader's convenience, we present here the structure of the paper.
In Section~\ref{sec:setup} we introduce our general setup and give a brief overview of already known solutions. 
The reader interested in the construction of scalar and vector oscillatons should read Section~\ref{sec:osci}. There, we first review how scalar oscillatons are constructed and then present for the first time details on massive vector field oscillatons.

Sections~\ref{sec:fluid} and~\ref{sec:collision} are devoted to the
study of stars with bosonic cores and their growth. We first study
stellar configurations formed by both a perfect fluid and a real massive
scalar field in Section~\ref{sec:fluid}. These solutions are a
generalization of the fluid-boson stars found and studied in detail in
Refs.~\cite{Henriques:1989ar,Henriques:1989ez,Lopes:1992np,Henriques:2003yr}. To
understand how these stars might grow, in Section~\ref{sec:collision} we
study head-on collisions of oscillatons in full Numerical Relativity. We
confirm and extend previous results, showing that collapse to a BH can
be avoided whenever gravitational cooling mechanisms are
efficient~\cite{Alcubierre:2003sx,Seidel:1993zk,Guzman:2006yc,Madarassy:2014jfa}. This
suggests that previous claims, assuming the collapse of the host star to
a BH above a certain threshold might not always be valid if DM is composed of massive bosonic fields.

Those interested in similar solutions in scalar-tensor theories should jump to Section~\ref{sec:ST}, where we consider perfect fluid stars in massless scalar-tensor theories and argue that stars with long-lived bosonic cores can exist in these theories.
We conclude our study in Section~\ref{sec:conclusion}, with possible extensions and follow-ups of this work.

\section{Setup}\label{sec:setup}

\subsection{Notation and conventions}

Unless otherwise and explicitly stated, we use geometrized units where
$G=c=1$, so that energy and time have units of length. 
We also adopt the $(-+++\dots)$ convention for the metric. For reference, the following is a list of
symbols that are used often throughout the text.

\begin{widetext}
\begin{table}[h]
\begin{tabular}{ll}
  $(t,r,\theta,\varphi)$ & Spherical coordinates employed in most of this work.\\
  $A_\mu$      & Electromagnetic four-vector potential, with mass parameter $\mu_V$.\\
  $A(\phi)$    & Generic function of a scalar field, in the context of scalar-tensor theories of gravity.\\
  $A_{\rm th}$ & Amplitude of thermal motion of nucleon inside star.\\
  $B$          & Metric component. We deal with diagonal metrics of the form $-F(t,r) dt^2+B(t,r)dr^2+r^2d\Omega^2$.  \\
  $l$          & Integer angular number, related to the eigenvalue $l(l+1)$ of spherical harmonics.\\ 
  $\lambda$    & Coupling constant for quartic interactions, described by a Lagragian ${\cal L}=-\lambda\phi^4/4$.\\
  $M_{\rm max}$& Peak value of the mass of a star in a mass-radius relation (cf. Fig.~\ref{MvsR}). Determines threshold for stability.\\
  $M_\odot$    & $\sim 1.989\times 10^{30}$ Kg, mass of our Sun.\\
  $M_T$        & Total ADM or gravitational mass, as measured by observers at large distances.\\
  $M_{F,B}$    & Time-average total mass in fluid and bosons, respectively.\\
  $M_0$        & Mass of standard star, in absence of scalar field, for the same value of central density.\\
  $m_N$        & Baryon mass.\\
  $\mu_{S,V}$  & mass parameter of (scalar or vector) bosonic fields. The physical mass $m_B$ is written as $m_B=\hbar \mu_{S,V} $.  \\
  $N_{F,B}$    & Total number of fermions and bosons in star, associated with conserved baryon number and Noether charge.\\
  $n_F$        & Baryonic number density in fluid frame.\\
  $\phi$       & Scalar field, taken to be fundamental and of mass $\mu_S$.\\ 
  $\rho_\phi$  & Energy density of scalar field.\\
  $\rho_F$     & Energy density, in the fluid frame, of a perfect fluid usually model with a polytropic equation of state $P=K\rho_F^{\gamma}$.\\
  $P$          & Pressure of perfect fluid. \\
  $R$          & Radius of bosonic condensate, defined to be the radius containing $98\%$ of the mass.\\
  $w_{1,2}$    & Width of scalar pulse initial data, in the context of collisions of scalar condensates.\\
  $\omega$     & Fourier transform variable, parameterizing the time dependence of variables, $\sim e^{-i\omega t}$.  \\
  $\Omega_{\rm th}$ & Vibration frequency of nucleon inside fluid star, due to thermal motion.\\ 
  $V$          & Radial, coordinate velocity of a given fluid element in the star. 
\end{tabular}
\end{table}
\end{widetext}

\clearpage
\newpage
\subsection{Framework}
We will be interested in a massive scalar $\phi$ or vector $A_{\mu}$ minimally coupled to gravity, and described by the action
\beq
S&=&\int d^4x \sqrt{-g} \left( \frac{R}{\kappa} - \frac{1}{4}F^{\mu\nu}\bar{F}_{\mu\nu}- \frac{\mu_V^2}{2}A_{\nu}\bar{A}^{\nu}\right.\nonumber\\
&-&\left.\frac{1}{2}g^{\mu\nu}\bar{\phi}^{}_{,\mu}\phi^{}_{,\nu} -\frac{\mu_S^2\bar{\phi}\phi}{2}
+\mathcal{L}_{\rm matter}\,
\right)
\,.\label{eq:MFaction}
\eeq
We take $\kappa=16\pi$, $F_{\mu\nu} \equiv \nabla_{\mu}A_{\nu} - \nabla_{\nu} A_{\mu}$ is the Maxwell tensor and $\mathcal{L}_{\rm matter}$ describes additional matter fields, that we consider to be described by a perfect fluid. We focus on massive, non self-interacting fields, but our results are easily generalized. In fact, we discuss in Section~\ref{sec:collision} how our results 
generalize to a quartic self-interaction term. The mass $m_B$ of the boson under consideration is related to the mass parameter above through $\mu_{S,V}=m_{B}/\hbar$, and the theory is controlled by the dimensionless coupling
\begin{equation}
\frac{G}{c\hbar} M_T\mu_{S,\,V} = 7.5\cdot 10^{9} \left(\frac{M_T}{M_{\odot}}\right) \left(\frac{m_{B}c^2}{eV}\right)\,,\label{dimensionless_massparameter}
\end{equation}
where $M_T$ is the total mass of the bosonic configuration.

Varying the action~\eqref{eq:MFaction}, the resulting equations of motion are
\begin{subequations}
\label{eq:MFEoMgen}
\begin{eqnarray}
  \label{eq:MFEoMScalar}
  &&\nabla_{\mu}\nabla^{\mu}\phi =\mu_S^2\phi
			\,,\\
  \label{eq:MFEoMVector}
  &&\nabla_{\mu} F^{\mu\nu} =
      \mu_V^2A^\nu\,,\\
  \label{eq:MFEoMTensor}
  &&\frac{1}{\kappa} \left(R^{\mu \nu} - \frac{1}{2}g^{\mu\nu}R\right)=
      \frac{1}{4\pi}\left(\frac{1}{2}F^{(\mu}_{\,\,\alpha}\bar{F}^{\nu)\alpha}- \frac{1}{8}\bar{F}^{\alpha\beta}F_{\alpha\beta}g^{\mu\nu}\right.\nonumber\\
   &&  \left. - \frac{1}{4}\mu_V^2A_{\alpha}\bar{A}^{\alpha}g^{\mu\nu}+\frac{\mu_V^2}{2}A^{(\mu}A^{\nu)} \right)
    -\frac{1}{4}g^{\mu\nu}\left( \bar{\phi}^{}_{,\alpha}\phi^{,\alpha}+{\mu_S^2}\bar{\phi} \phi\right)   \nonumber\\
   &&   +\frac{1}{4}\bar{\phi}^{,\mu}\phi^{,\nu}+\frac{1}{4}\phi^{,\mu}\bar{\phi}^{,\nu}+\frac{1}{2}T_{\rm fluid}^{\mu\nu}\,.
%
\end{eqnarray}
\end{subequations}
Here, the stress-energy tensor for the perfect fluid is given by~\cite{rezzolla2013relativistic}
\be\label{stress_energy_PF}
T_{\rm fluid}^{\mu\nu}=\left(\rho_F+P\right)u^{\mu}u^{\nu}+P g^{\mu\nu}\,,
\ee
with $u^{\mu}$ the fluid's four-velocity, $\rho_F$ its total energy density in the fluid frame and $P$ its pressure.
The massive vector field equations~\eqref{eq:MFEoMVector} imply that the vector field must satisfy the constraint,
\be\label{divA}
\mu^2_V\nabla_{\mu}A^{\mu}=0\,,
\ee
while from the Bianchi identities, it follows that the fluid must satisfy the conservation equations
\be\label{divT}
\nabla_{\mu}T_{\rm fluid}^{\mu\nu}=0\,.
\ee
In addition we impose conservation of the baryonic number~\cite{rezzolla2013relativistic}: 
\be\label{baryonic}
\nabla_{\mu}\left(n_F u^{\mu}\right)=0\,,
\ee
where $n_F$ is the baryonic number density in the fluid frame and $m_N n_F$ is the fluid's rest-mass density for baryons of mass $m_N$.
To close this system of equations we also need to complement the system with an equation of state relating $n_F$, $\rho_F$ and $P$. Specific equations of state will be discussed in Sec.~\ref{sec:fluid}.

In the following we consider only everywhere regular solutions of the system~\eqref{eq:MFEoMScalar}--\eqref{eq:MFEoMTensor}.
\subsection{Brief overview of solutions}

Compact solutions of the system~\eqref{eq:MFEoMScalar}--\eqref{eq:MFEoMTensor}, without including the perfect fluid, exist for both real and complex fields. Scalar fields have been extensively studied in the literature, while similar solutions for vector fields have only recently been shown to exist~\cite{Garfinkle:2003jf,Brito:2015yga,Brito:2015pxa}. Here we give a brief summary of the most popular solutions which have been studied so far. For more detailed reviews on the subject see Refs.~\cite{Jetzer:1991jr,Schunck:2003kk,Liebling:2012fv,Macedo:2013jja}.

\paragraph{Boson stars.}
Boson stars are regular compact solutions of the Einstein-Klein-Gordon equations for a {\it complex} massive scalar field. 
Some of these solutions have been claimed in the literature as possible candidates for supermassive horizonless BH mimickers~\cite{Torres:2000dw}.
They can be classified according to the scalar potential in the Klein-Gordon Lagrangian~\eqref{eq:MFaction}~\cite{Schunck:2003kk}:
\begin{itemize}
\item {Mini boson stars}: the scalar field potential is given by $V(\phi)=\mu_S^2|\phi |^2$, where $\mu_S$ is the scalar field mass. 
For non-rotating boson stars the maximum mass is $M_{{\rm max}}\approx 0.633 m_{\rm P}^2/\mu_S$, with $m_{\rm P}$ being the Planck mass~\cite{Kaup:1968zz,Ruffini:1969qy}. 
Considering values of $\mu_S$ typically found within the Standard Model, this mass limit is much smaller than the Chandrasekhar limit for a fermion star, approximately $m_{\rm P}^3/\mu^2$. For ultralight boson masses $\mu_S$, as those motivated by string axiverse scenarios~\cite{Arvanitaki:2009fg}, and relevant in the DM context, mini boson stars may have a total mass compatible with that observed in active galactic nuclei~\cite{Schunck:2003kk}.
\item {Massive boson stars}: the scalar potential has an additional quartic scalar field term, $V(\phi)=\mu_S^2|\phi|^2+\lambda |\phi|^4/2$~\cite{Colpi:1986ye} (see also Ref.~\cite{Eby:2015hsq} for a detailed study of these solutions in the context of DM physics). In this case the maximum mass can be comparable to the Chandrasekhar limit and for $\lambda \gg \mu_S^2/m_{\rm P}^2$ one can estimate $M_{{\rm max}}\approx 0.062\lambda^{1/2}m_{\rm P}^3/\mu_S^2$. 
\end{itemize}
One can also find other types of boson stars by changing the scalar potential or by considering non-minimal couplings as done in Ref.~\cite{vanderBij:1987gi}. For some non-linear potentials, similar solutions, generically called Q-balls, exist even in flat space~\cite{Coleman:1985ki,Lee:1991ax} (when coupled to gravity some of these solutions can give rise to very heavy boson stars~\cite{Friedberg:1986tq}. See also Ref.~\cite{Schunck:1999zu} for potentials with more generic self-interaction terms). Finally, boson stars can also be found in alternative theories of gravity, such as scalar-tensor theories~\cite{Torres:1997np,Torres:1998xw,Whinnett:1999ma}. A more detailed list of solutions can be found in Ref.~\cite{Schunck:2003kk}.

\paragraph{Oscillatons.}

For a {\it real} massive scalar field minimally coupled to gravity, with $V(\phi)=\mu_S^2\phi^2$, compact configurations were first shown to exist in Ref.~\cite{Seidel:1991zh}, while the generalization for a scalar field with a quartic interaction was considered in~\cite{UrenaLopez:2012zz}. Interestingly, for some non-linear potentials, such as the Higgs double well potential or the axionic sine-gordon potential, a real scalar field counterpart of the Q-balls exist and are dubbed oscillons~\cite{Bogolyubsky:1976nx,Gleiser:1993pt,Copeland:1995fq,Kolb:1993hw,Gleiser:2006te} (note that oscillons are built in a {\it Minkowski} background).  
Unlike boson stars, since the scalar field stress-energy tensor is itself time-dependent, for oscillatons both the metric and the scalar field oscillate periodically in time. Oscillatons (and oscillons) are not truly periodic solutions of the field equations, as they decay through quantum and classical processes. However, their lifetime $T_{\rm decay}$ is extremely large for the masses of interest~\cite{Page:2003rd,Grandclement:2011wz},
\be
T_{\rm decay}\sim 10^{324}\left(\frac{1\,{\rm meV}}{m_B{\rm c^2}}\right)^{11}\, {\rm yr}\,.
\ee
For real massive vector fields, Ref~\cite{Garfinkle:2003jf} found convincing indications that the same kind of oscillatory solutions form in the gravitational collapse of a wide set of initial data.

\begin{figure}[htb]
\begin{center}
\begin{tabular}{c}
\epsfig{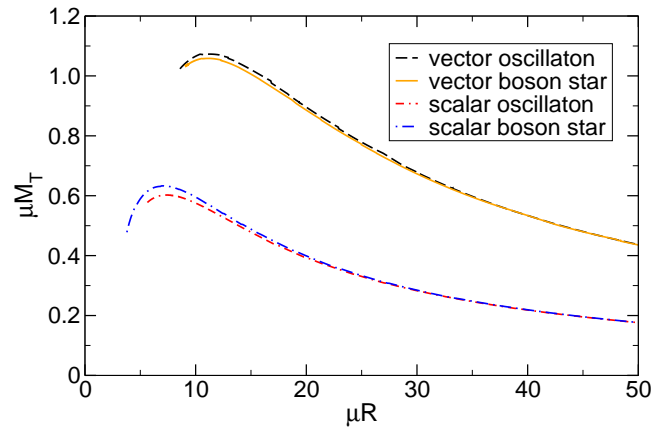}
\end{tabular}
\caption{Comparison between the total mass of a boson star ({\it complex} scalar or vector fields) and an oscillaton ({\it real} scalar or vector fields), as a function of their radius $R$. $R$ is defined as the radius containing 98\% of the total mass. The procedure to find the diagram is outlined in the main text.\label{MvsR}}
\end{center}
\end{figure}

Boson stars and oscillatons share very similar structures, as summarized in Fig.~\ref{MvsR}, where we plot the mass-radius relation for spherically symmetric boson stars and oscillatons (including massive vectors that will be discussed in the next Section). Boson stars and oscillatons have a maximum mass $M_{\rm max}$, given approximately by
\be
\frac{M_{\rm max}}{M_{\odot}}=8\times 10^{-11}\,\left(\frac{\rm eV}{m_{B}c^2}\right)\,,\label{max_mass}
\ee
for scalars and slightly larger for vectors. 

\paragraph{Boson-Fermion stars}

The extension to mixed stars, composed both by a complex scalar field and a perfect fluid, was first considered in Ref.~\cite{Henriques:1989ar} and further studied in~\cite{Henriques:1989ez,Lopes:1992np,Henriques:2003yr} (exact solutions in $2+1$-dimensions were also found in~\cite{Sakamoto:1998aj}). The stability of these objects was studied in~\cite{Jetzer:1990xa,Henriques:1990xg,Henriques:1990vx,ValdezAlvarado:2012xc}. Slowly rotating boson-fermion stars were constructed in~\cite{deSousa:2000eq}, while extensions to allow for an interaction between the scalar field and the fermionic fluid were considered in~\cite{deSousa:1995ye,Pisano:1995yk}.

Boson-fermion stars were shown to exist in a wide variety of configurations. For small boson masses they can be either dominated by the bosonic component or the fermionic, or have bosonic and fermionic components of the same order of magnitude~\cite{Henriques:1989ar,Henriques:1989ez,Lopes:1992np,Henriques:2003yr}. For large boson masses only two types of configurations exist, either bosonic dominated or fermionic dominated, with a sharp transition between the two configurations, when one of the configurations reaches $\sim 10\%$ of the total mass~\cite{Henriques:1989ez,Lopes:1992np}.

In this paper we will show that similar solutions also exist for real massive fields. However in this case the fluid must itself oscillate with a dominant frequency given by twice the boson mass. 

%

\section{Oscillatons}\label{sec:osci}

In this Section we construct compact solutions of the Einstein field equations, either for a minimally coupled \emph{real} massive scalar or vector field. We first review the formalism introduced in Ref.~\cite{Seidel:1991zh} to construct massive scalar oscillatons and then show how this formalism can be extended to massive vector fields.   

\subsection{Massive scalar field}

We start by considering a \emph{real} massive scalar minimally coupled to gravity. In Ref.~\cite{Seidel:1991zh} it was shown that solutions to the field equations describing spherically symmetric compact configurations exist. We consider a general time-dependent spherically symmetric metric
\be\label{metric}
ds^2=-F(t,r)dt^2+B(t,r)dr^2+r^2d\Omega^2\,.
\ee
Rescaling the scalar field as $\phi\to \phi/\sqrt{8\pi}$ and defining the function $C(t,r)=B(t,r)/F(t,r)$, the field equations~\eqref{eq:MFEoMScalar} and~\eqref{eq:MFEoMTensor} lead to a system of partial differential equations (PDEs) given by
\beq\label{eqs_scalar}
\frac{\dot{B}}{B}&=&r \dot{\phi}\phi'\,,\label{eqs_scalar1}\\
\frac{B'}{B}&=&\frac{r}{2}\left(C\dot{\phi}^2+(\phi')^2+B\mu_S^2\phi^2\right)+\frac{1}{r}\left(1-B\right)\label{eqs_scalar2}\,,\\
\frac{C'}{C}&=&\frac{2}{r}\left[1+B\left(\frac{1}{2}r^2\mu_S^2\phi^2-1\right)\right]\,,\label{eqs_scalar3}\\
\ddot{\phi}C&=&-\frac{1}{2}\dot{C}\dot{\phi}+\phi''+\phi'\left(\frac{2}{r}-\frac{C'}{2C}\right)-B\mu_S^2\phi\label{eqs_scalar4}\,,
\eeq
where an overdot denotes $\partial/\partial t$ and a prime denotes $\partial/\partial r$.
These equations suggest the following periodic expansion
\beq\label{series_scalar}
B(t,r)&=&\sum_{j=0}^{\infty} B_{2j}(r)\,\cos\left(2j\omega t\right)\,,\nonumber\\
C(t,r)&=&\sum_{j=0}^{\infty} C_{2j}(r)\,\cos\left(2j\omega t\right)\,,\nonumber\\
\phi(t,r)&=&\sum_{j=0}^{\infty} \phi_{2j+1}(r)\,\cos\left[\left(2j+1\right)\omega t\right]\,.
\eeq
Inserting this expansion into the system~\eqref{eqs_scalar2}--\eqref{eqs_scalar4} and truncating the series at a given $j$, yields a set of ordinary differential equations for the radial Fourier components of the metric functions and the scalar field. We note that out of the four  Eqs.~\eqref{eqs_scalar1}--~\eqref{eqs_scalar4} we only need to use three. The remaining one can be checked to be satisfied a posteriori. In practice we only compute the Fourier expansion~\eqref{series_scalar} up to $j=j_{\rm max}$. This introduces a certain error in the accuracy at which the full system of equations is satisfied. In general the larger $j_{\rm max}$ is, the smaller the error~\cite{Fodor:2009kg,Grandclement:2011wz}, and we explicitly checked that this was the case. 

We impose regular boundary conditions at $r=0$, i.e., $\phi_{2j+1}'(0)=0$,  $B_{0}(0)=1$, $B_{2j}(r)=0$ for $j\geq 1$, while $\phi_{2j+1}(0)$ and $C_{2j}(0)$ are free parameters. At infinity $r\to\infty$, asymptotic flatness requires $\phi_{2j+1}\to 0$, $C_{0}=B_{0}\to 1$ and $C_{2j}=B_{2j}\to 0$ for $j\geq1$\footnote{For asymptotically flat spactimes the metric function $C(t,r)\to B^2(r)/\alpha(t)$ when $r\to\infty$, for some arbitrary function $\alpha(t)$. Thus we can always rescale $t$ such that $C(t,r)\to B^2(r)$ at $r\to \infty$.}. 
The system~\eqref{eqs_scalar2}--\eqref{eqs_scalar4} supplemented with this set of boundary conditions is an eigenvalue problem for the frequency $\omega$. Fixing one of the free parameters, e.g, $\phi_{1}(0)$, one can shoot for the other remaining free parameters, requiring that the boundary conditions are satisfied. For each choice of $\phi_{1}(0)$ there will be a unique family of solutions satisfying the above boundary conditions, characterized by the number of nodes in the scalar field profile.

Due to the presence of a mass term in the scalar potential, the scalar field decays in a Yukawa-like fashion $e^{-r\sqrt{\mu_S^2-\omega^2}}/r$ at large distances. Thus, at infinity the metric asymptotically approaches the Schwarzschild solution and the total mass of a given configuration can be computed using
\be\label{ADM_mass}
M_T=\lim_{r\to\infty} m(r)=\lim_{r\to\infty} \frac{(B-1)r}{2B}\,.
\ee
We should note that oscillatons are not truly stable configurations, but decay on very long time-scales due to a radiative tail. 
However, since the amplitude of this tail is exponentially suppressed, for our purposes it is enough to compute the mass at some finite radius and consider it to be the mass of the oscillaton.
Although these stars do not possess a well-defined surface where the field vanishes, the configuration is exponentially suppressed at a radius $r\sim 1/\mu_S$. Thus, one can define an effective radius inside which much of the mass is localized. We will define the radius $R$ of the oscillaton as being the radius such that $m(R)$ is $98\%$ of the total mass $M_T$.  

\begin{figure*}[htb]
\begin{center}
\begin{tabular}{cc}
\epsfig{file=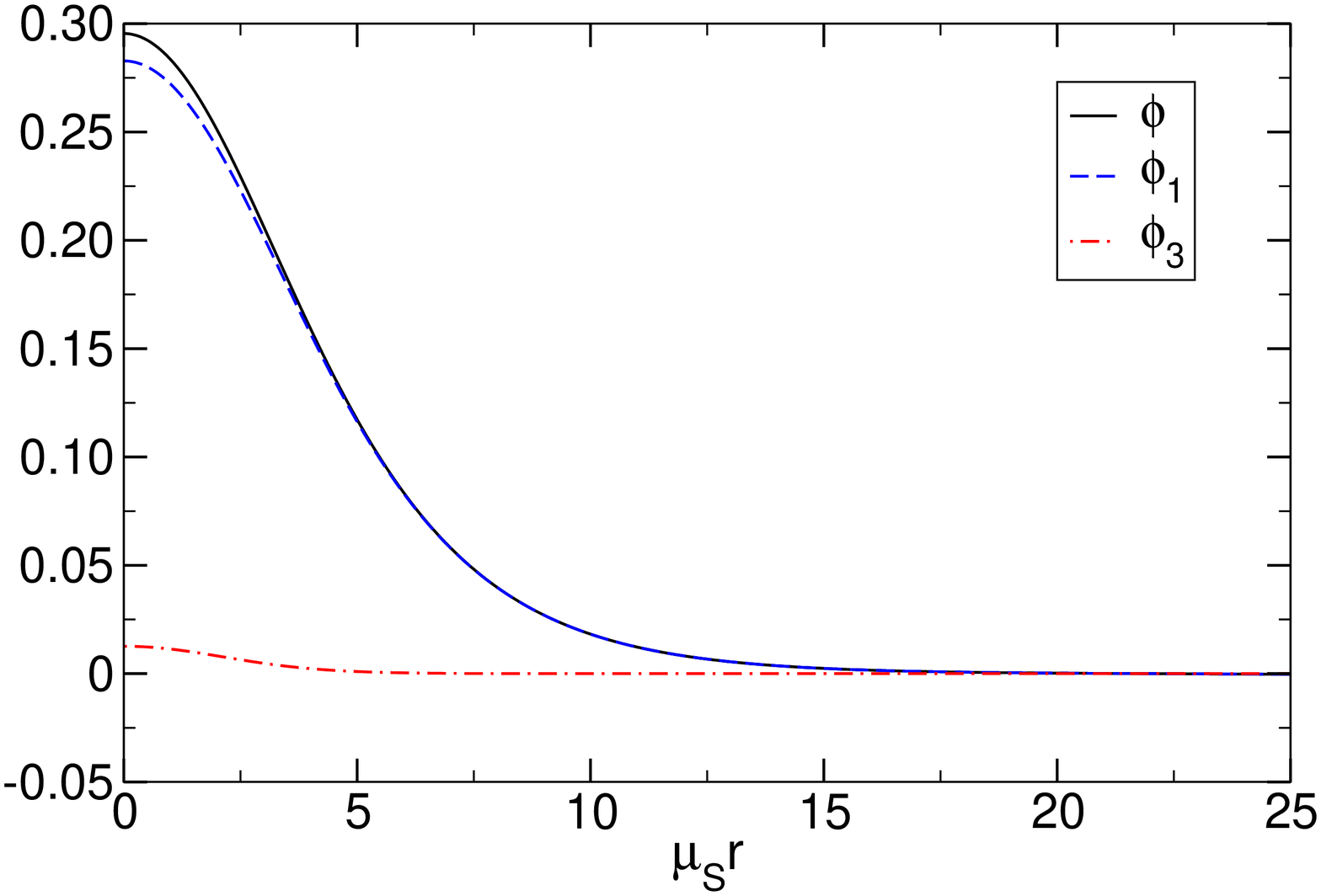,width=6.9cm,angle=0,clip=true}\hspace{6em}
\epsfig{file=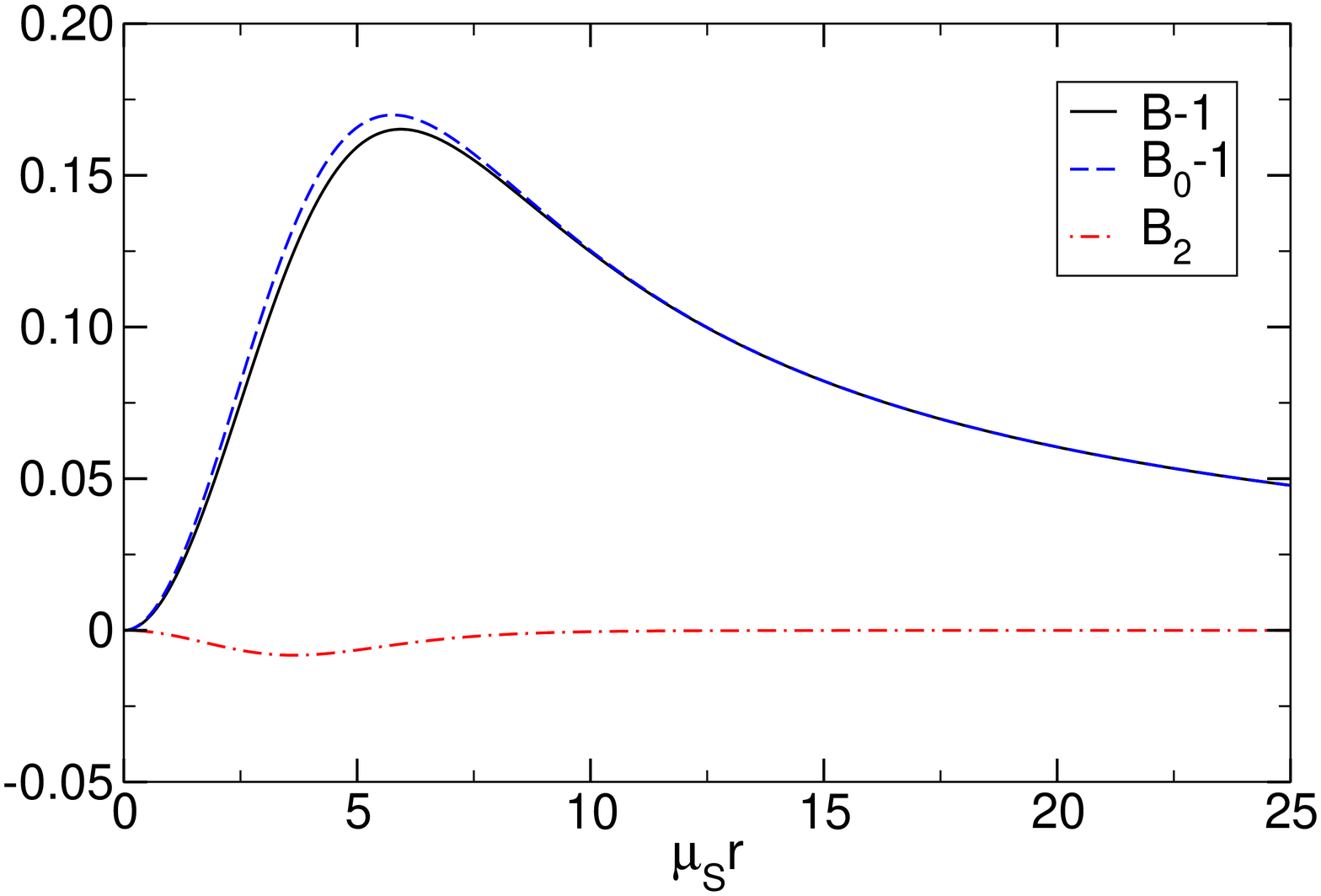,width=6.9cm,angle=0,clip=true}
\end{tabular}
\caption{Left: Scalar field configuration $\phi$ at $\omega t=0$ for $\phi_1(0)=0.2828$ and its first Fourier components with $j_{\rm max}=1$. Right: Corresponding radial metric coefficient $B$ at $\omega t=0$ and its first Fourier components. This configuration has a total mass $M_T\approx 0.57/\mu_S$ and fundamental frequency $\omega\approx 0.912\mu_S$.\label{scalar_osci}}
\end{center}
\end{figure*}

\begin{figure}[htb]
\begin{center}
\begin{tabular}{c}
\epsfig{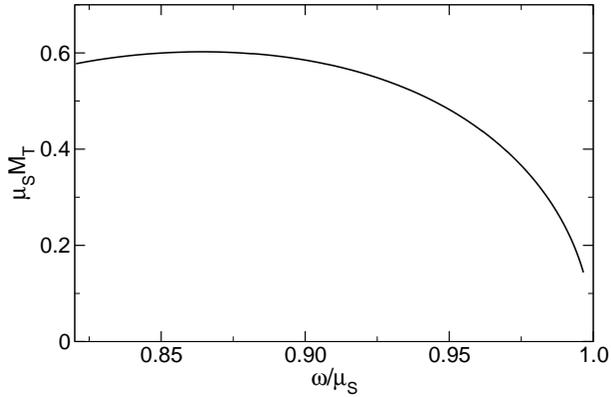}
\end{tabular}
\caption{Total mass $M_T$ of the scalar oscillaton as a function of the fundamental frequency $\omega$. The maximum mass is $M_T\sim 0.6/ \mu_S$ for $\omega \sim 0.864\mu_S$. This point marks the threshold between stable and unstable configurations. Stars to the right of the maximum are stable while those to the left are unstable.\label{scalar_Mvsw}}
\end{center}
\end{figure}

In Fig.~\ref{scalar_osci} we show an example of a configuration. The profile is smooth everywhere and we find that the series~\eqref{series_scalar} typically converges already for $j=2$. The fundamental frequency satisfies $\omega\lesssim \mu_S$ as shown in Fig.~\ref{scalar_Mvsw}, where we plot the mass $M_T$ as a function of $\omega$. In the Newtonian limit $M_T\to 0$, the solutions become spatially diluted with $\omega\to \mu_S$ (cf. Fig.~\ref{MvsR}). For smaller $\omega$ the star becomes more compact, with a maximum mass given by $M_T\sim 0.6/\mu_S$ for $\omega \sim 0.864\mu_S$, in agreement with previous studies~\cite{Seidel:1991zh,Alcubierre:2003sx}. 

Due to the time-dependence of these solutions, a perturbative analysis of their linear stability is very challenging. However the close similarity between oscillatons and boson stars, suggest that the solutions are stable from $\omega=\mu_S$ down to the maximal mass~\cite{Gleiser:1988ih,Lee:1988av}. The results of Refs.~\cite{Seidel:1991zh,Alcubierre:2003sx}, where Numerical Relativity techniques were used to study how oscillatons behave when slightly perturbed, suggest that this is indeed the case. Since scalar oscillatons have been widely discussed in the literature, we will not discuss them further and instead show that similar configurations exist for massive vector fields. 

%

\subsection{Massive vector field}

Very recently, the Einstein-Proca field equations~\eqref{eq:MFEoMVector} and~\eqref{eq:MFEoMTensor} have been shown to admit compact configurations, similar to scalar boson stars, for a complex massive vector field~\cite{Brito:2015pxa}. On the other hand, for real vector fields, Ref.~\cite{Garfinkle:2003jf} found strong numerical evidences that vector oscillatons can form in the gravitational collapse of a real massive vector field. Using the formalism introduced above, we can show that real massive vector fields can indeed form oscillating compact structures when minimally coupled to gravity. This is, as far as we are aware, the first time that these solutions are explicitly constructed. 

We consider the metric~\eqref{metric} and a spherically symmetric vector field
\be
A_{\mu}dx^{\mu}=A_t(t,r) dt+A_r(t,r) dr\,.
\ee
To simplify the equations we define the following functions
\be
X_t\equiv \sqrt{C} A_t\,,\quad
X_r\equiv \frac{1}{\sqrt{C}} A_r\,,\quad
W\equiv \frac{\sqrt{C}}{B}\left(\dot{A}_r-A'_t\right),
\ee
where again $C\equiv B/F$.
From eqs.~\eqref{eq:MFEoMVector} and~\eqref{divA} we find
\begin{align}
&X'_t=\frac{X_r\dot{C}}{2}+C\dot{X}_r-\frac{X_t}{r}\left(B-1\right)
-BW\left(1-r W X_t\right)\,,\\
&\dot{X}_t=\frac{1}{r^2}\partial_r\left(r^2 X_r\right)\label{EQMax_2}\,,\\
&X_t=-\frac{1}{\mu_V^2r^2}\partial_r\left(r^2W\right)\label{EQMax_3}\,,\\
&\dot{W}=-\mu_V^2 X_r\label{algebraic_W}\,,
\end{align}
while from Einstein's field equations~\eqref{eq:MFEoMTensor} we have
\beq
\frac{\dot{B}}{B}&=&2 r \mu^2_V X_r X_t \label{eq_vector1}\,,\\
\frac{B'}{B}&=&r\left(\mu_V^2 C X_r^2+\mu_V^2 X_t^2+BW^2\right)+\frac{1}{r}\left(1-B\right)\,,\\
\frac{C'}{C}&=&\frac{2}{r}\left[1+B\left(r^2W^2-1\right)\right]\,.
\eeq
These equations suggest the following Fourier expansions
\beq\label{series_vector}
X_t(t,r)&=&\sum_{j=0}^{\infty} X_{t\,2j+1}(r)\,\cos\left[\left(2j+1\right)\omega t\right]\,,\nonumber\\
X_r(t,r)&=&\sum_{j=0}^{\infty} X_{r\,2j+1}(r)\,\sin\left[\left(2j+1\right)\omega t\right]\,,\nonumber\\
W(t,r)&=&\sum_{j=0}^{\infty} W_{2j+1}(r)\,\cos\left[\left(2j+1\right)\omega t\right]\,,
\eeq
while the metric functions are expanded as in~\eqref{series_scalar}. Once more, eq.~\eqref{eq_vector1} will not be used to find the solutions. On the other hand, from Eq.~\eqref{algebraic_W}, one can find $W_{2j+1}$ algebraically, which greatly simplifies the equations (note that $W$ is just an auxiliary function that we introduced to simplify the equations, and so one can easily check that, after finding $W_{2j+1}$ from eq.~\eqref{algebraic_W} and $X'_{r\,2j+1}$ from eq.~\eqref{EQMax_2}, eq.~\eqref{EQMax_3} is automatically satisfied).
Similarly to the scalar case, we can find a set of ordinary differential equations by truncating the series at some $j$ and then solve the eigenvalue problem, imposing regular boundary conditions at $r=0$ and asymptotic flatness. This imposes $X_{r\, 2j+1}(0)=0$,  $B_{0}(0)=1$, $B_{2j}(r)=0$ for $j\geq 1$, while $X_{t\, 2j+1}(0)$ and $C_{2j}(0)$ are free parameters. At infinity, besides the usual conditions for the metric functions, we require $X_{t\,2j+1}=X_{r\,2j+1}\to 0$. We can then find solutions by fixing $X_{t\,1}(0)$ and use the same method as for the scalar case.

\begin{figure*}[htb]
\begin{tabular}{cc}
\epsfig{file=vector_A,width=6.9cm,angle=0,clip=true}\hspace{6em}
\epsfig{file=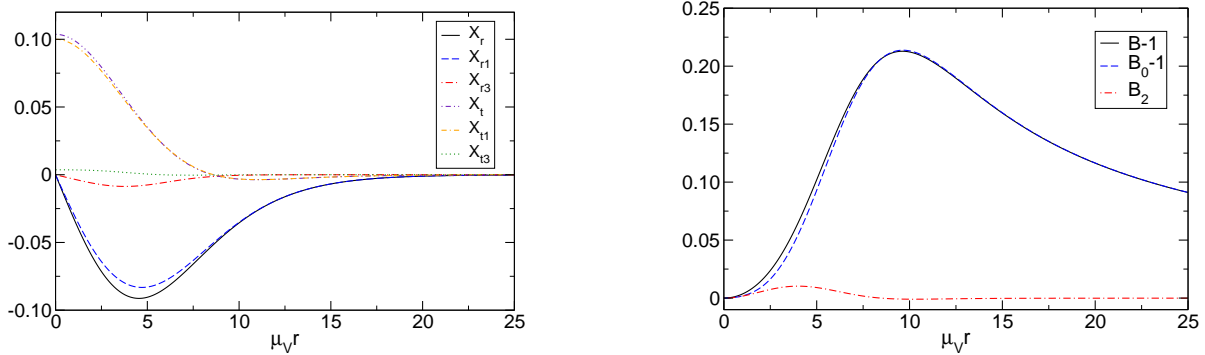,width=6.9cm,angle=0,clip=true}
\end{tabular}
\caption{Left: Vector field profiles at their peak values for $X_{t\,1}(0)=0.1$ and its first Fourier components with $j_{\rm max}=1$. Right: Corresponding radial metric coefficient $B$ at $\omega t=0$ and its first Fourier components. This configuration has a total mass $M_T\approx 1.044/\mu_V$ and fundamental frequency $\omega\approx 0.902\mu_V$.\label{vector_osci}}
\end{figure*}
\begin{figure}[htb]
\begin{center}
\begin{tabular}{c}
\epsfig{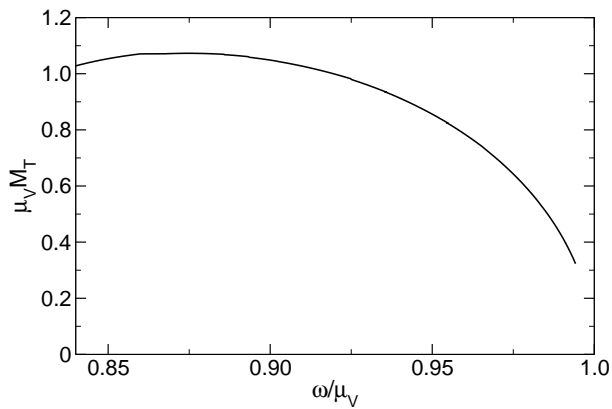}
\end{tabular}
\caption{Total mass $M_T$ of the vector oscillaton as a function of the fundamental frequency $\omega$. The maximum mass is $M_T\approx 1.07/ \mu_V$ for $\omega \approx 0.875\mu_V$. This point marks the threshold between stable and unstable configurations.\label{vector_Mvsw}}
\end{center}
\end{figure}

Our results are summarized in Figs.~\ref{vector_osci}--\ref{vector_Mvsw}. The overall behavior is analogous to the scalar case. 
The series~\eqref{series_vector} converges rapidly and already for $j=2$ one gets an accuracy for the ADM mass better than $\sim 0.2\%$. The total mass $M_T$ as a function of frequency $\omega$ is shown in Fig.~\ref{vector_Mvsw}. The behavior is analogous to the one found in the scalar case, although the maximum is slightly larger, $M_T\sim 1.07/ \mu_V$ (cf. Fig~\ref{MvsR}), and occurs at $\omega \sim 0.875\mu_V$. Not surprisingly, the overall behavior is almost identical to the one found for Proca stars (i.e. complex vector field boson stars)~\cite{Brito:2015pxa}, as can be seen in Fig.~\ref{MvsR}. 

By considering radial perturbations of Proca stars, it was shown in Ref.~\cite{Brito:2015pxa} that the maximum mass also corresponds to a branching point separating unstable from stable solutions. Although full numerical simulations are needed, vector oscillatons should also follow the same pattern. In particular, similar conclusions should hold: configurations which reach the unstable branch will either quickly collapse to BHs or migrate back to the stable branch via mass ejection, a phenomenon known as the
gravitational cooling mechanism~\cite{Alcubierre:2003sx,Seidel:1993zk,Guzman:2006yc} (see also Section~\ref{sec:collision}).

As a final word, we should note that although we only discussed fundamental states, characterized by $X_t$ having one node and $X_r$ being nodeless, excited states~--~solutions with more nodes~--~also exist. Since those are expected to be unstable~\cite{Balakrishna:1997ej,Balakrishna:2007mr} we will not discuss them here.

%

\section{Stars with dark matter cores}\label{sec:fluid}
\subsection{Setting and Fourier-expansion}\label{subsec:setting}
In the previous Section we showed that self-gravitating massive bosonic fields can form compact configurations. 
An interesting possibility is that these structures might be accreted by stars, forming a bosonic core in their interior. For complex bosonic fields, such solutions were already considered in the literature~\cite{Henriques:1989ar,Henriques:1989ez,Lopes:1992np,Henriques:2003yr,Sakamoto:1998aj,deSousa:2000eq,deSousa:1995ye,Pisano:1995yk}. In a recent \emph{Letter}~\cite{Brito:2015yga} we showed that real bosonic fields can also cluster inside stars and give rise to oscillating configurations, where both the star's material and the field oscillate. In the following we provide details on these solutions.

We consider the system~\eqref{eq:MFEoMScalar},~\eqref{eq:MFEoMTensor}, and for simplicity we focus only on the scalar case. Similar configurations should also exist for vector fields.
Due to the presence of the scalar field the fluid will, in general, oscillate with a radial velocity:
\be
dr/dt=u^r/u^0=V(t,r)\,,
\ee
where $u^0=\Gamma/\sqrt{-g_{tt}}$ and $\Gamma=\left(1-U^2\right)^{1/2}$ is the Lorentz factor connecting the fluid's comoving frame with the frame of the observer at rest with respect to a spacelike hypersurface of constant $t$~\cite{1991A&A...252..651G}. Using the normalization condition $u^{\mu}u_{\mu}=-1$ one has $U=V\sqrt{-g_{rr}/g_{tt}}$.
In the \emph{Letter}~\cite{Brito:2015yga} we did not assume conservation of the baryonic number, given by Eq.~\eqref{baryonic}. This is equivalent to neglecting the radial velocity $V\sim 0$. As we show below, this is indeed a good approximation for small bosonic cores. Physically, non conservation of the baryonic number leads to a conversion of bosonic matter to baryonic matter and vice-versa. This mimics a fundamental interaction between the scalar field and star's material for which the baryon number is not conserved. On the other hand, whether a conversion between fundamental fields can indeed occur indirectly through their gravitational coupling is an interesting discussion that we leave for future work. We should also note that for \emph{complex} fields, which give rise to the boson-fermion stars first studied in Refs.~\cite{Henriques:1989ar,Henriques:1989ez}, the fluid and the metric are static, and so for this case we set $V=0$.

Once more, we will work with a rescaled scalar field, $\phi\to \phi/\sqrt{8\pi}$ and consider a general time-dependent, spherically symmetric metric, as in Eq.~\eqref{metric}.
The field equations, together with the conservation equations~\eqref{divT} and~\eqref{baryonic} lead to a system of PDEs given by
\beq
&&\dot{B}/B=r\left[\dot{\phi}\phi'-\frac{8\pi B V(P+\rho_F)}{1-U^2}\right]\,,\label{cont_eq1}\\
&&\dot{\rho}_{F}=\left(P+\rho_F\right)\frac{\dot{n}_B+Vn_F'}{n_F}-V\rho'_{F} \,,\label{cont_eq12}\\
&&B'/B=\frac{r}{2}\left[C\dot{\phi}^2+(\phi')^2+B\left(\mu_S^2\phi^2+16\pi\frac{\rho_{F}+PU^2}{1-U^2}\right)\right]\nn\\
&&+(1-B)/r\,,\\
&&C'/C=2/r+Br\left(\mu_S^2\phi^2+8\pi\rho_{F}-8\pi P\right)-2B/r\,,\\
&&\ddot{\phi}C=-\dot{C}\dot{\phi}/2+\phi''+2\phi'/r-C'\phi'/(2C)-B\mu_S^2\phi\,,\\
&&2P'=-\left(1-U^2\right)\left(P+\rho_{F}\right)\left(CB'-BC'\right)/(BC)\nn\\
&&+V\left[\left(P+\rho_F\right)\left(4CV-r\dot{C}\right)-2rC\dot{P}\right]/r\nn\\
&&-2\left(P+\rho_{F}\right)C\dot{V}+2CV\left(P+\rho_F\right)\frac{\dot{n}_B+Vn_F'}{n_F}\,,\\
&&V'=-\left(1-U^2\right)\left[\left(\dot{B}+VB'\right)/(2B)+\left(\dot{n}_B+Vn_F'\right)/n_F\right]\nn\\
&&-V\left[4+C\left(2r\dot{V}-4V^2\right)+rV\left(\dot{C}+VC'\right)\right]/(2r)
\,.\label{cont_eq2}
\eeq
where we recall that $C\equiv B/F$ and $U\equiv\sqrt{C}V$. We will not make use of the conservation equations~\eqref{cont_eq1} and ~\eqref{cont_eq12}. One can check a posteriori that these equations are satisfied up to a certain error introduced by the ansatz we use.
%
%
%
%
%
%

Employing the periodic expansion~\eqref{series_scalar}, one can easily see that the fluid's energy density, rest-mass density, pressure and radial velocity can be consistently expanded as 
\beq\label{series_fluid}
\rho_F(t,r)&=&\sum_{j=0}^{\infty} \rho_{F\,2j}(r)\,\cos\left(2j\omega t\right)\,,\nonumber\\
n_F(t,r)&=&\sum_{j=0}^{\infty} n_{F\,2j}(r)\,\cos\left(2j\omega t\right)\,,\nonumber\\
P(t,r)&=&\sum_{j=0}^{\infty} P_{2j}(r)\,\cos\left(2j\omega t\right)\,,\nonumber\\
V(t,r)&=&\sum_{j=1}^{\infty} V_{{2j}}(r)\,\sin\left(2j\omega t\right)\,,
\eeq
The equations of motion need to be supplemented by an equation of
state. We will focus on an ideal fluid and polytropic equation of state~\cite{rezzolla2013relativistic}:
\be
P=K \left(m_N n_F\right)^{\gamma},\quad \rho_F(P)=\left(P/K\right)^{1/\gamma}+P/\left(\gamma-1\right)\,,
\ee
where we take $K=100/\mu_S^2$ and $\gamma=2$, which can mimic neutron stars~\cite{1965ApJ...142.1541T,ValdezAlvarado:2012xc}. For this choice, the star is also isentropic, i.e. the fluid's specific entropy is constant along the star~\cite{rezzolla2013relativistic}.
In the \emph{Letter}~\cite{Brito:2015yga} we considered the equation of state $P=K \rho_F^{\gamma}$, which is equivalent to the previous one when the fluid's internal energy density is much smaller than the fluid's rest-mass density. This is a good model for cold and old neutron stars~\cite{rezzolla2013relativistic}. In the following, we will compare the results obtained in both models.
In geometrical units $G=c=1$ and $\gamma=2$, $\sqrt{K}$ has units of length and can be used to set the length-scale of the problem in the absence of the scalar field~\cite{1964ApJ...140..434T}. Without loss of generality we will also set $m_N=1$. The choice $K=100$ was considered in e.g. Ref.~\cite{ValdezAlvarado:2012xc}, which we used to check the accuracy of our code. Other values of $K$ can be obtained by fixing $\mu_S$ and rescale all the quantities accordingly. For example, fixing $\mu_S=1$, mass and radius are measured in units of $\sqrt{K}$.
Although our results can be generalized to other equations of state, we should note that generic equations of state do not allow for a straightforward expansion such as~\eqref{series_fluid}. A possibility is to consider the oscillating components to be a small perturbation of a static star, along the lines of what is usually done to construct slowly-rotating stars~\cite{Hartle:1967he,Hartle:1968si}~\footnote{In this case, at lowest order, the expansion~\eqref{series_fluid} would be given by $\rho_F(t,r)=\rho_{F\,0}(r)+\epsilon^2 P_{2}(r)\,\partial\rho_{F\,0}/\partial P_{0}\cos\left(2\omega t\right)$, $n_F(t,r)=n_{F\,0}(r)+\epsilon^2 P_{2}(r)\,\partial n_{F\,0}/\partial P_{0}\cos\left(2\omega t\right)$, $P(t,r)=P_{0}(r)+\epsilon^2 P_{2}(r)\cos\left(2\omega t\right)$, with $\epsilon$ a small bookkeeping parameter and we assume an equation of state in the absence of the scalar field of the form $\rho_{F\,0}\equiv \rho_{F\,0}(P_{0})$ and $n_{F\,0}=n_{F\,0}(P_{0})$.}. The construction here presented can then be straightforwardly applied. We have explicitly checked that using this approach one can generalize our results to a generic equation of state.

\subsection{Numerical procedure}
To construct the stars we employ the same method used in the previous Section, the difference being that, due to the presence of the fluid, the solutions are now parametrized by 
two parameters, e.g., $n_{F\,0}(0)$ and $\phi_1(0)$, while for the radial velocity we impose $V_{2j}(0)=0$. Additionally, we also need to impose boundary conditions at the star's radius. We define the radius $R$ of the star to be the location where the pressure drops to zero, $P(R)=0$.
For high scalar field central densities, first-order terms $j=1$ in the density might become of the order of the zeroth-order term, making it difficult to find these configurations with good accuracy and impose the boundary condition at the star's radius. However, as explained below, for a given $n_{F\,0}(0)$, we expect these configurations to become unstable at some threshold $\phi_1(0)>\phi^c_{1}(0)$. To avoid these numerical difficulties, we will mostly focus on small $\phi_1(0)$.
  
Due to the different length scales present in the problem, the solutions can also be characterized by the mass coupling $\mu_S M_0$, where $M_0$ is the mass of the \emph{static} star for vanishing scalar field, corresponding to the same value of central rest-mass density $n_{F\,0}(0)$. Depending on the numerical value of $\mu_S M_0$, we employ different numerical strategies. For small $\mu_S M_0$, the scalar field density profile extends beyond the star's radius. For this case we compute the profile inside the star and at the star's radius impose the matching with the outer solution. The full solution is then found by imposing asymptotically flat boundary conditions. For large $\mu_S M_0$, the scalar field is exponentially suppressed inside the star. To avoid numerical errors to spoil the full solution, we perform tree integrations: we first find the radius at which the scalar field drops to zero and then compute the remaining solution by imposing the scalar field to be zero after this radius. The solution outside the star is then found by matching it with the inner solution.

A useful quantity to describe scalar-fluid stars is the scalar field's energy density, given by
\be\label{scalar_density}
2\rho_{\phi}=-2T_0^{\phantom{0}0}=-\dot{\phi}^{2}/g_{tt}+\phi^{'2}/g_{rr}+\mu_S^2\phi^2\,,
\ee
and the energy density measured by an observer at rest with respect to a spacelike hypersurface of constant $t$, given by
\be\label{fluid_density}
\rho_{\mathcal{F}}=-T_0^{\phantom{0}0}=\Gamma^2\left(\rho_{F}+P\right)-P\,.
\ee
Note that, for our solutions, the contribution from the fluid's kinetic energy to $\rho_{\mathcal{F}}$ is negligible, and so we have in general $\Gamma\sim 1$ and $\rho_{\mathcal{F}}\sim\rho_F$.
With this, we define the time-average total mass in the fluid and bosons as
\be
M_{F,\,B}=\int_0^{\infty} 4\pi \left<\sqrt{B}\,\rho_{{\mathcal{F}},\,\phi}\right> r^2 dr\,,\label{fermion_mass}
\ee
where $<>$ denotes a temporal average. The total mass $M_T$ can be found in the usual way through the metric component $g_{rr}$ which asymptotically approaches the Schwarzschild solution at infinity (cf. Eq.~\eqref{ADM_mass}).

\begin{figure}[ht]
\begin{center}
\begin{tabular}{c}
\epsfig{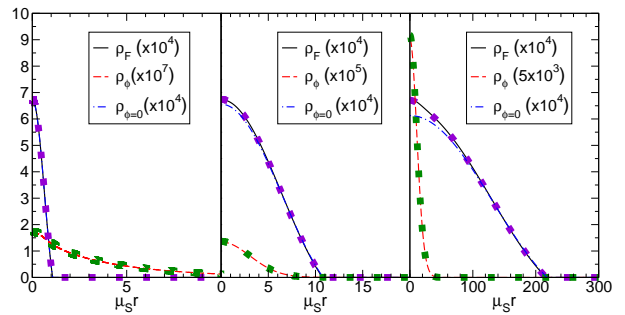}
\end{tabular}
\caption{
Comparison between the (time average) energy density of the scalar field $\rho_{\phi}$ and the fluid $\rho_{F}$ for mixed scalar oscillatons and fermion fluids, for scenarios where the total baryon number is conserved (fluid velocity $V\neq 0$). 
We fix $\rho_{F\,0}(0)=0.0006332$, and have
from left to right, $\mu_S M_0=0.1,\,M_B/M_T\approx 21\%$, corresponding to $\phi_1(0)=0.026$ and $\omega M_0\approx 0.0993$,
$\mu_S M_0=1,\,M_B/M_T\approx 0.66\%$, corresponding to $\phi_1(0)=0.025$ and $\omega M_0\approx 0.863$,
and $\mu_S M_0=20,\,M_B/M_T\approx 0.54\%$, corresponding to $\phi_1(0)=0.015$ and $\omega M_0\approx 16.221$.
Squares denote the corresponding quantities for complex fields (i.e. mixed boson-fluid stars for the same $M_F$ and $M_B$). The overlap is nearly complete.
Here $M_0$ and $\rho_{\phi=0}$ are the total mass of the star (for the same $\rho_{F\,0}(0)$) and the energy density of the fluid (for the same $M_F$), respectively, when the scalar field vanishes everywhere. In the left panel, the $\rho_{\phi=0}$ and the $\rho_{F}$ lines are indistinguishable, because light fields have a negligible influence on the fluid distribution. For $\rho_{F\,0}(0)=0.0006332$, we have, in our units, $M_0=1$ which corresponds to a star with $M_0\sim M_{\odot}$. Solutions with larger $\rho_{F\,0}(0)$ can also be obtained and the qualitative picture remains the same. 
\label{scalarfluid_osci}}
\end{center}
\end{figure}
%
\subsection{Results}
We will discuss our results assuming baryon conservation during DM accretion and dynamics,
but we will also discuss stars for which the DM-baryon cross section is so large that
conversion between one and the other is extreme, to the point where solutions with zero fluid velocity
are allowed. These solutions conserve baryon number on the average, but not instantaneously.
Additionally, this also serves as a model for stars composed of fields for which there is no conserved current, such as Majorana fermions or real bosonic fields.
\subsubsection{Conserved baryon number}
%
\begin{figure*}[ht]
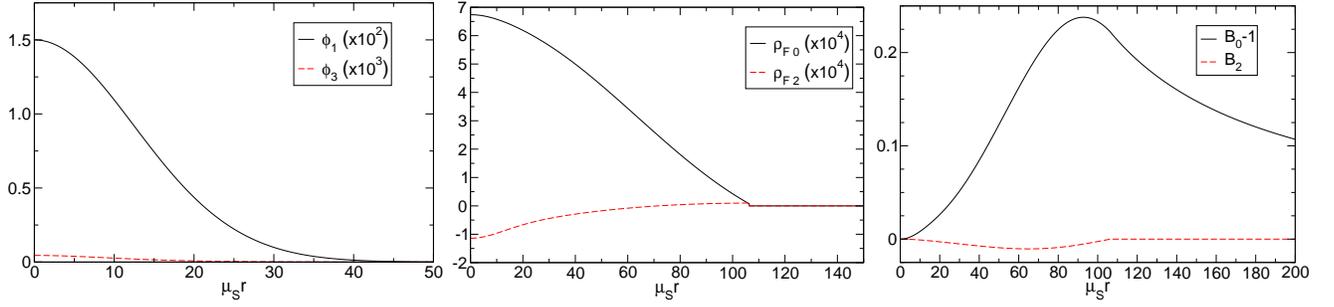

\begin{center}
\begin{tabular}{ccc}
\epsfig{file=scalarfluid_phi_mu10_MBperc06,width=5.8cm,angle=0,clip=true}
\epsfig{file=scalarfluid_rho_mu10_MBperc06,width=5.5cm,angle=0,clip=true}
\epsfig{file=scalarfluid_grr_mu10_MBperc06,width=5.8cm,angle=0,clip=true}
\end{tabular}
\caption{Scalar field (left) configuration, density profile of the fluid (center), and corresponding metric component $g_{rr}$ (right) for $\mu_S M_0=10$ and $M_B/M_T\approx 0.54\%$, corresponding to $\phi_1(0)=0.015$, $n_{F\,0}(0)=0.0006332$ and $\omega M_0\approx 8.118$. We plot the first Fourier components for $j_{\rm max}=1$.\label{scalarfluid_solution_v2}}
\end{center}
\end{figure*}
\begin{figure}[htb]
\begin{center}
\begin{tabular}{c}
\epsfig{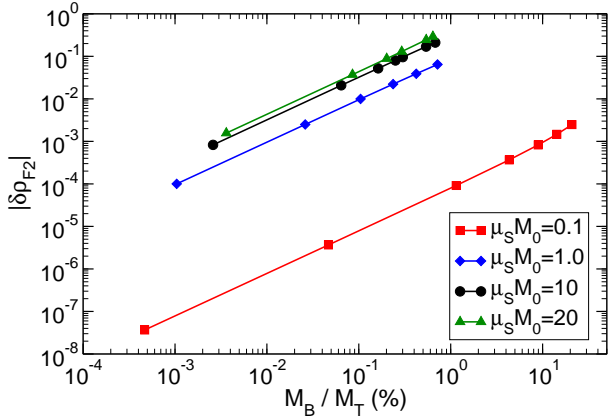}
\end{tabular}
\caption{
Amplitude of the oscillations $\delta\rho_{F\,2}\equiv\rho_{F\,2}(0)/\rho_{F\,0}(0)$ as a function of $M_B/M_T$. The symbols denote actual solutions that we computed. We find that for small mass ratios $M_B/M_T$ and large $\mu M_0$ the amplitude is well fitted by $\delta\rho_{F\,2}\sim 10(\mu M_0)^{1/2} M_B/M_T$.
\label{osci_linear}}
\end{center}
\end{figure}

\begin{figure}[htb]
\begin{center}
\begin{tabular}{cc}
\epsfig{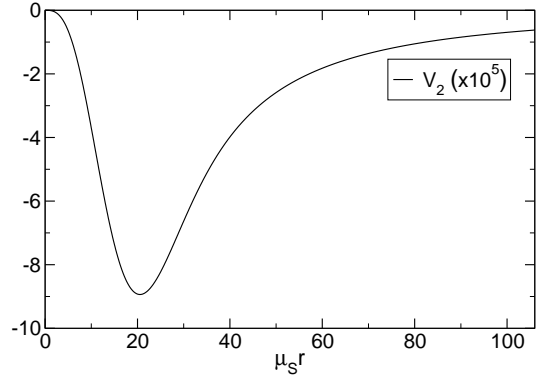}
\end{tabular}
\caption{Velocity profile of the fluid for the star in Fig.~\ref{scalarfluid_solution_v2} ($V_2$ is the dominant time-dependent component of the velocity profile, cf. equation~\eqref{series_fluid}).\label{scalarfluid_solution_v2_vel}}
\end{center}
\end{figure}

Our results for composite stars with conserved total baryon numbers are summarized in Figs.~\ref{scalarfluid_osci}-\ref{scalarfluid_solution_v2_vel}. The overall behavior and global structure of these DM-cored stars is dependent on the new mass scale introduced by the scalar field mass. For very small $\mu M_0$, the Compton wavelength of the scalar is very large and the scalar field spreads throughout the spacetime. This is shown in the left panel of Fig.~\ref{scalarfluid_osci} for $\mu_SM_0=0.1$.
For very large scalar field masses, on the other hand, the scalar is confined to a small region inside the star, as seen in the right panel of Fig.~\ref{scalarfluid_osci} for $\mu_SM_0=20$. In fact, when $\mu_S M_0$ is extremely large, as happens for many DM models (c.f. \eqref{dimensionless_massparameter}), the scalar core hardly knows about the existence of the star outside, and behaves, to a very good precision, exactly like the pure oscillatons we described in the last Sections.
Notice also that for large mass couplings, one can have a large density, small oscillaton inside a fluid star.
As we discuss below, our argument then indicates that the oscillaton can be in the stable branch, indicating that the whole configuration is stable.
In other words, stable, self-gravitating bosonic DM cores inside stars are possible. These results complement similar recent findings for fermionic DM cores~\cite{Leung:2013pra}.

Accordingly, the detailed structure of these stars will also depend on the dimensionless coupling $\mu_S M_0$.
For large couplings, one can think of these composite stars as a regular fluid star, where at the center sits a small
pulsating oscillaton. It is then natural to expect that the oscillations in the oscillaton density will {\it induce oscillations} in the fluid material. Indeed, this is a generic feature borne out of our results.
A typical star structure is shown in Fig.~\ref{scalarfluid_solution_v2} for a 0.54\% scalar composition
and $\mu M_0=10$, corresponding to a scalar core well inside the fermion fluid. A general feature of these stars is
that they {\it oscillate}, driven by the scalar field, with a frequency
\be
f=2.5\times 10^{14}\,\,\frac{m_{B}c^2}{\rm eV}\,{\rm Hz}\,.
\ee
In particular, the local density is a periodic function of time with a period dictated by the scalar field. These oscillations are signalled by a nonzero $j=1$ component of the fermion density expansion \eqref{series_fluid}, and are driven by the time-varying component of the oscillaton's density; therefore, the {\it amplitude} of the fluid's oscillations is expected to scale with the mass of the oscillaton. As shown in Fig.~\ref{osci_linear}, for small mass ratios $M_B/M_T$ and large $\mu M_0$ we find that the amplitude of these oscillations is described by the approximate relation
\be
 \delta\rho_{F\,2}\equiv\rho_{F\,2}(0)/\rho_{F\,0}(0)\sim 10 (\mu M_0)^{1/2} M_B/M_T\,.
\ee
Even for $M_B/M_T=0.01$, and for $\mu_S M_0=10$ the oscillations are of the order of $30\%$ of the static component.
For this particular setup, where the baryon number is conserved, the fluid velocity is nonzero, and is shown in Fig.~\ref{scalarfluid_solution_v2_vel}.

\begin{figure}[htb]
\begin{center}
\begin{tabular}{c}
\epsfig{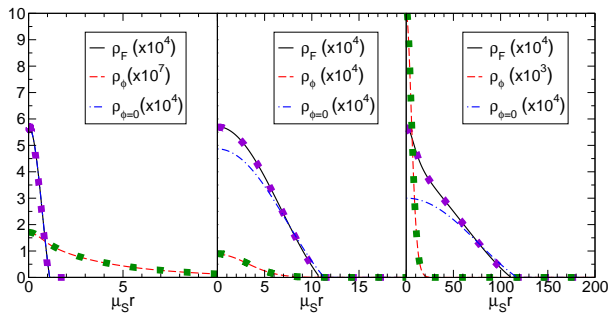}
\end{tabular}
\caption{Same as Fig.~\ref{scalarfluid_osci} but for the case considered in the $\emph{Letter}$, i.e., $V=0$ and $P=K\rho_F^{\gamma}$.
We fix $\rho_{F\,0}(0)=0.00057092$, and have from left to right, $\mu_S M_0=0.1,\,M_B/M_T\approx 21\%$, corresponding to $\phi_1(0)=0.026$ and $\omega M_0\approx 0.0993$,
$\mu_S M_0=1,\,M_B/M_T\approx 5\%$, corresponding to $\phi_1(0)=0.064$ and $\omega M_0\approx 0.873$,
and $\mu_S M_0=10,\,M_B/M_T\approx 5\%$, corresponding to $\phi_1(0)=0.06982$ and $\omega M_0\approx 8.629$.
Once more, squares denote the corresponding quantities for complex fields (i.e. mixed boson-fluid stars for the same $M_F$ and $M_B$).
\label{scalarfluid_osci_2}}
\end{center}
\end{figure}
%
\subsubsection{Non-conservation of baryon number}
%
\begin{figure*}[htb]
\begin{center}
\begin{tabular}{ccc}
\epsfig{file=scalarfluid_phi_mu10_MBperc5,width=5.5cm,angle=0,clip=true}
\epsfig{file=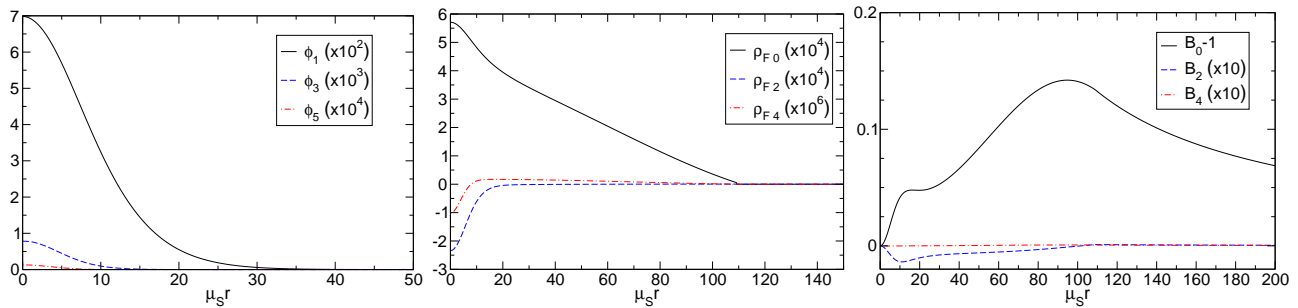,width=5.5cm,angle=0,clip=true}
\epsfig{file=scalarfluid_grr_mu10_MBperc5,width=5.8cm,angle=0,clip=true}
\end{tabular}
\caption{Scalar field (left) configuration, density profile of the fluid (center) and corresponding metric component $g_{rr}$ (right) for $\mu_S M_0=10$ and $M_B/M_T\approx 5\%$, corresponding to $\phi_1(0)=0.06982$, $\rho_{F\,0}(0)=0.00057092$ and $\omega M_0\approx 8.629$. Here we take $V=0$ and consider $P=K\rho_F^{\Gamma}$. We plot the first Fourier components for $j_{\rm max}=2$.
\label{scalarfluid_solution}}
\end{center}
\end{figure*}
\begin{figure}[htb]
\begin{center}
\begin{tabular}{cc}
\epsfig{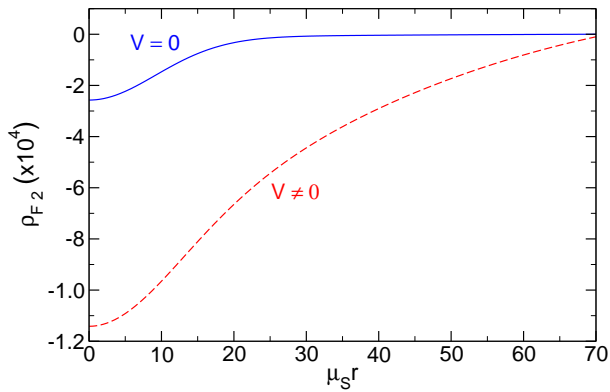}
\end{tabular}
\caption{For the sake of comparison, we show here the amplitude of the oscillations ($\rho_{F2}$ is the dominant time-dependent component of the fluid density, cf. equation (\ref{series_fluid})) for $\mu_S M_0=10$ and $M_B/M_T\approx 0.54\%$, corresponding to the solution of Fig.~\ref{scalarfluid_solution_v2}, computed using $V=0$ with the equation of state $P=K\rho_F^{\gamma}$ and $V\neq 0$ with the equation of state $P=K \left(m_Nn_F\right)^{\gamma}$, $\rho_F(P)=\left(P/K\right)^{1/\gamma}+P/\left(\gamma-1\right)$.\label{scalar_fluid_comparison}}
\end{center}
\end{figure}
The previous results can be compared and contrasted with the extreme case where DM and baryonic matter can convert into one another.
Under this assumption, we find that there are solutions, summarized in Figs.~\ref{scalarfluid_osci_2}-\ref{scalar_fluid_comparison}, that allow a star to have zero fluid velocity (see also discussion in Section~\ref{subsec:setting}). We find that despite this, the overall qualitative behavior is the same as those of baryon-conserving stars.

The density distribution of baryon non-conserving stars is shown in Fig.~\ref{scalarfluid_osci_2} for different dimensionless mass couplings. We remind the reader that these are $V=0$ and $P=K\rho_F^{\gamma}$ stars that were considered in the \emph{Letter}. 
Again, the mass coupling changes drastically the global behavior of these stars; large mass couplings result in a small bosonic DM core
which is oblivious of the fermions surrounding it.

The structure of a star is shown in Fig.~\ref{scalarfluid_solution}. Because the equations are technically less challenging to handle in this case, we can accurately compute their $j=2$ Fourier components and consider larger values of $M_B/M_T$.
As we said, the qualitative behavior is similar, and in particular these composite stars also oscillate in density, driven by the density-varying oscillaton sitting at their center.
Since the effect of the radial velocity is negligible for small $M_B/M_T$, taking $V\sim 0$ describes with very good accuracy the main properties of these stars. For large $\mu M_0$, the only noticeable effect of $V$ on the solution is to slightly increase the amplitude of the oscillations in comparison to the $V=0$ case. A comparison for the solution of Fig.~\ref{scalarfluid_solution_v2}, where this difference is noticeable, is shown in Fig.~\ref{scalar_fluid_comparison}. In general, the smaller $\mu M_0$ and $M_B/M_T$ the better the agreement between the two cases.

We stress that the overall behavior might have been anticipated from an analysis of Fig.~\ref{MvsR}: for light fields, $\mu_S M_0<1$ the scalar profile is extended and the pure oscillaton solution is broad and light. As such, the scalar has a negligible influence on the fluid distribution (as can be seen from the fact that the zero-scalar line $\rho_{\phi=0}$ overlaps with the fluid line in Figs.~\ref{scalarfluid_osci} and~\ref{scalarfluid_osci_2}), and these stars simply have an extended scalar condensate protruding away from them.
In fact, our results are compatible with a decoupling between the boson and fluid for large $\mu_S M_B$. For this case, Fig.~\ref{MvsR} alone is enough to interpret the bosonic distribution.
For example, for $\mu_S M_0=10, M_B/M_T=5\%$, we get $\mu_S M_B\sim 0.3$, which would imply from Fig.~\ref{MvsR} that $\mu_S R\sim 20$ for the scalar field distribution. This is indeed apparent from Fig.~\ref{scalarfluid_osci_2}.
Similar conclusions were reached when studying mixed fermion fluid/boson stars with complex fields~\cite{Lopes:1992np}. In fact, the structure of mixed oscillatons and fluid stars is almost identical to that of boson stars and fluids, as can be seen from Figs.~\ref{scalarfluid_osci} and~\ref{scalarfluid_osci_2}, where we overplot with dotted lines the complex field case.

Overall, our results are consistent with what was previously found for boson-fermion fluid stars~\cite{Henriques:1989ar,Henriques:1989ez}. We expect that field configurations with high $M_B/M_T$ for large $\mu_S M_0$ should follow the same kind of behavior as that found in boson-fermion stars. In particular we expect that bosonic dominated stars should also be possible when increasing $M_B/M_T$~\cite{Henriques:1989ar,Henriques:1989ez}.

Finally, as we discuss below, a careful stability analysis shows that, for sufficiently small $\phi_1(0)$ and for stars which are stable in the absence of scalars,
composite stars are dynamically stable. On the other hand, our results show that these configurations can be understood well from the mass-radius relation of oscillatons.
The maximum mass supported is (\ref{max_mass}), $M_{\rm max}/M_{\odot}=8\times 10^{-11}\,eV/(m_{B}c^2)$, which for a neutron star and an axion field of mass $10^{-5}\,eV$ falls well within the stability regime~\cite{Gleiser:1988}.

\subsubsection{Stability of fluid-boson stars}

As was mentioned before, the stability properties of fluid-boson stars are not expected to depend on the details of the scalar field description. Therefore, we
will focus here on the well-studied case of a fermionic star with a complex scalar field, and assume that the results will hold in more generic cases (in particular, when the scalar field is real).
Since fermion-boson star solutions depend on two parameters (i.e., for instance $\{n_{F}(0), \phi_1(0)\}$), the stability theorems for single parameter solutions can not be directly applied. This implies that the change in stability of these solutions cannot easily be inferred from the extremes of a mass versus radius diagram, and requires a more careful analysis. Nevertheless, one can argue that a {\it necessary} condition for stability is that the binding energy $M_T-m_N N_F-m_B N_B$ be negative~\cite{Henriques:1989ez}, where $N_B$ is the number of bosons
(associated to the conservation of the Noether charge) and $N_F$ is the number of fermions
(associated to the conservation of the baryonic number) defined, respectively, by
\beq
N_{F}=\int_0^{\infty} 4\pi \sqrt{B}\,n_{F} r^2 dr\,,\\
N_{B}=\int_0^{\infty} 4\pi \sqrt{C}\,\omega|\phi|^2 r^2 dr\,.
\eeq

For the solutions shown in Fig.~\ref{osci_linear}, the binding energy defined in terms of the masses $M_T-M_F-M_B$ is always negative. Although negative values do not necessarily imply stability, they do give strong support to the claim that these configurations are stable. A more careful stability analysis shows that for sufficiently small $\phi_1(0)$ and for stars which are stable in the absence of scalars, the negativity of the binding energy is a good criterion for stability~\cite{Henriques:1990xg}.

A more strict stability criterion to find the critical point --which separates the stable from the unstable configurations-- can be obtained through a dynamical analysis, since at the critical point the lowest eigenvalue in the (radial) perturbations of the star passes zero. In general this dynamical analysis might become rather involved and simpler alternative approaches have been proposed. For instance, it has been observed that in a critical point there must be a direction $\mathbf{n}$ such that the directional derivatives of $\{M,N_F,N_B\}$ vanish~\cite{Henriques:1990xg}, implying that at the stability boundary $\mathbf{n}$ is tangential to the level curves of constant $M$, $N_B$, and $N_F$. Using this property,  the stability boundary can be found by drawing contours onto the plane $\{n_F(0), \phi_1(0)\}$ for fixed values of the particle numbers, and looking for the points where these curves meet and are tangential to each other~\cite{Henriques:1990xg} (see e.g. Fig. 1 of Ref.~\cite{Henriques:1990xg} and Fig. 2 of Ref.~\cite{ValdezAlvarado:2012xc}). 

It was also noticed~\cite{ValdezAlvarado:2012xc} that, since a level curve of constant total mass $M_c$ implicitly defines the trajectory $\phi_1(0) = \phi_1(0)[n_F(0),M_c]$, the equilibrium critical configurations also correspond to the extreme values of the number of particles when surveyed along a level curve of constant total mass. Therefore, a simple procedure to find the critical point is to calculate the extremes of the particle numbers when the total mass is held fixed:
\beq
\frac{\partial N_F}{\partial n_F(0)} \bigg|_{M_T=M_c} = \frac{\partial N_B}{\partial n_F(0)}\bigg|_{M_T=M_c} = 0   ~~~.
\eeq
This method was employed in~\cite{ValdezAlvarado:2012xc} to obtain the critical point and distinguish stable from unstable configurations. These results were validated by numerical evolutions, showing that only the stars on one side of such extreme were stable against perturbations. The stars on the other side of the critical point were unstable and could, depending on the initial perturbation, either migrate to a stable star or collapse to a BH. Unfortunately, only cases with a small dark matter component were considered, so the question regarding what happens with self-gravitating scalar fields remained unanswered.

These studies can be easily extended to fermion-boson stars with self-gravitating scalar fields by allowing for solutions with comparable number of bosonic and fermionic particles. Similar masses can be achieved by either modifying the EoS of the baryonic fluid or the mass of the boson particle. Here we allow for mixed stars dominated either by fermions or bosons by setting $\{K,m_B\}$ such that the maximum mass of isolated fermion and boson stars are the same $M_B=M_F=0.633$. 

\begin{figure}[htb]
\begin{center}
\begin{tabular}{cc}
 \epsfig{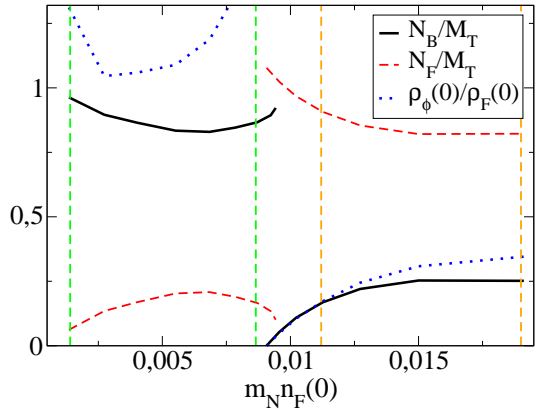}
\end{tabular}
\caption{Number of bosonic and fermionic particles as a function
of $n_F$, rescaled with the total mass $M_T=0.55$. The maximum mass for either an isolated boson star or fermion star is $M_{max}=0.633$. 
There are two solution regimes, one with $N_B>N_F$ for low fluid
densities and other with $N_F>N_B$ for high fluid densities. We
also display the ratio of energy densities at the origin $\rho_{\phi}(0)/\rho_F(0)$,
showing that the solutions with low fluid density correspond
to a self-gravitating scalar field. Each regime of equilibrium configurations
has a maximum/minimum corresponding to a critical point. Vertical 
lines represents the configurations, on both sides of each extreme,
being evolved in time to confirm their stability.
\label{particle_rho}}
\end{center}
\end{figure}

\begin{figure}[htb]
\begin{center}
\begin{tabular}{cc}
\epsfig{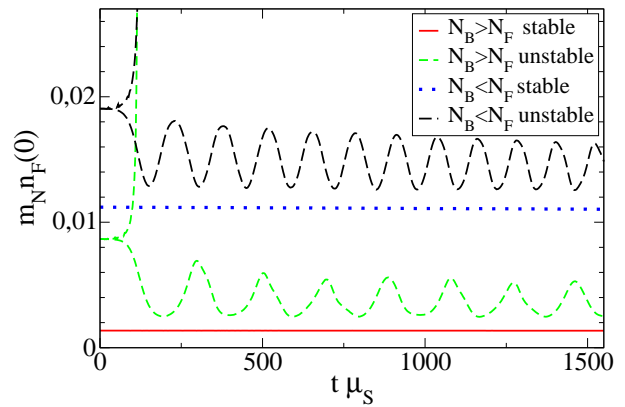}
\end{tabular}
\caption{Central fluid density $m_N n_F(0)$ as a function of time for
four representative equilibrium configurations from
fig.~\ref{particle_rho}. The two configurations on the left side of the local extremes are stable and remain constant (or oscillating) under small perturbations. The configurations on the right side
are unstable, and small perturbations lead the star either to migrate to a stable star or to collapse to a BH. 
\label{fb_evolution}}
\end{center}
\end{figure}

In Fig.~\ref{particle_rho} the number of particles $N_F$ and $N_B$, rescaled with the total constant mass $M_T=0.55$, are plotted as a function of the central rest-mass fluid density $m_N n_F(0)$. Clearly, there are two different solution regimes: at low fluid density the solutions are dominated by the scalar field component such that $N_B > N_F$ and 
$\rho_{\phi}(0) > \rho_{F}(0)$. On the other hand, for high fluid densities, the solutions are dominated by the fermionic part, so $N_F > N_B$ and
$\rho_{\phi}(0) < \rho_{F}(0)$. An extreme, corresponding to the critical point separating stable from unstable configurations, is present in these two regimes. It is easy to show that, in each regime, the branch on the left of the maximum correspond to the stable solutions. This result is indeed confirmed by numerical evolutions, which are performed with the code and the techniques described in detail in Ref.~\cite{ValdezAlvarado:2012xc}. Four representative equilibrium configurations, one on each side of the extreme present in each
regime, are modified by a small initial perturbation. The time evolution
of the central fluid density $m_N n_F(0)$ is displayed in Fig.~\ref{fb_evolution}. Under small perturbations, the stable configurations just oscillate with the quasi-normal modes, while unstable configurations either migrate to a stable star or collapse to a BH.

These results confirm that there is nothing special regarding fermion-boson stars with self-gravitating scalar fields, in contrast to some claims in the literature~\cite{Goldman:1989nd,Kouvaris:2011fi,Bramante:2014zca,Bramante:2015cua,Kurita:2015vga}; these configurations can be either stable or unstable, depending on the parameters of the system.

\section{Stars with scalar cores in scalar-tensor theories}\label{sec:ST}
Scalar fields are a fundamental component of scalar-tensor theories of gravity, one of the most natural and extensively studied extensions of General Relativity~\cite{Fujii:2003pa}. These theories are normally characterized by a non-minimal coupling between a scalar field and gravity. Very compact stars with vanishing scalar in scalar-tensor theories can be unstable towards the growth of a scalar field, a phenomena called {\it spontaneous scalarization}~\cite{Damour:1993hw,Pani:2010vc,Palenzuela:2013hsa,Berti:2015itd}. The final state is a static star with a non-vanishing but static scalar profile.

A natural consequence of our work is that in these theories stars can also have scalar cores if the scalar field is massive and time-dependent. Given that massive scalar-tensor theories are poorly constrained, this opens the way to improve current bounds on these theories from binary pulsar experiments. We will leave this for future work and focus instead on massless scalar-tensor theories. We argue that even for a massless scalar field, some theories might admit stars with long-lived scalar cores.

Focus on the simplest possible case, that of a complex scalar-tensor theory. This is conceptually easier to handle because it allows for the existence of spherically symmetric solutions with a static metric. This theory is formally equivalent to a tensor-multi-scalar theory with two real scalar fields~\cite{Damour:1992we,Horbatsch:2015bua}. 
Our results also apply to single scalar-tensor theories with nonminimally coupled real scalar fields, the difference being that in these theories the geometry must also oscillate. 

In the physical (Jordan) frame the scalar is non-minimally coupled to the Ricci scalar~\cite{Fujii:2003pa}. By performing a conformal transformation, one can write the theory in the Einstein frame as~\cite{Damour:1992we,Horbatsch:2015bua}
\beq\label{ST_action}
S&=&\int d^4 x\sqrt{-g}\left[\frac{R}{16\pi}-g^{\mu\nu}\partial_{\mu}\bar{\phi}\partial_{\nu}\phi
\right]\nn\\
&+&S_m\left[A^2\left(\phi,\bar{\phi}\right)g_{\mu\nu};\Phi\right]\,,
\eeq
where $\bar{\phi}$ denotes the complex conjugate of $\phi$, $A^2\left(\phi,\bar{\phi}\right)$ is a generic function of the scalar field, and $S_m$ denotes the matter action. The matter fields, denoted collectively by $\Phi$, are minimally coupled to the Jordan frame metric $\tilde{g}_{\mu\nu}=A^2\left(\phi,\bar{\phi}\right)g_{\mu\nu}$, where the tilde denote quantities computed in the Jordan frame. This guarantees that the weak equivalence principle holds. 
By varying the action~\eqref{ST_action}, one obtains the following scalar field equation (apart from the Einstein-Klein-Gordon equations minimally coupled to the matter fields):
\be
\Box \phi=
-2\frac{\partial \log A}{\partial\bar{\phi}} T\,,\label{ST_scalar}
\ee
where $T$ denotes the trace of the matter fields' stress-energy tensor. The physical stress-energy tensor (written in the Jordan frame) is related to the Einstein-frame stress energy tensor by
\be
T_{\nu}^{\mu}=A^4\tilde{T}_{\nu}^{\mu},\quad T_{\mu\nu}=A^2\tilde{T}_{\mu\nu}, \quad T=A^4\tilde{T}\,,
\ee
where $\tilde{T}^{\mu\nu}$ is the physical stress-tensor in the Jordan frame, given by Eq.~\eqref{stress_energy_PF} (for details see e.g. Ref.~\cite{Pani:2014jra}).

We will assume that at spatial infinity the scalar field vanishes and that the function $A$ can be expanded as
\be
A\approx 1+\alpha\phi+\bar{\alpha}\bar{\phi}+\frac{1}{2}\beta\phi\bar{\phi}+\frac{1}{4}\beta_1\phi^2+\frac{1}{4}\bar{\beta}_1\bar{\phi}^2+\ldots\,,
\ee
where $\beta$ is a real constant, while $\alpha$ and $\beta_1$ are complex numbers. Without loss of generality we set $\alpha=\beta_1=0$. 
Applying this expansion to Eq.~\eqref{ST_scalar} one immediately sees that the field acquires an effective position-dependent mass term given by $\mu_{\rm eff}^2=-\beta T$~\cite{Cardoso:2013fwa,Cardoso:2011xi}.
By taking the ansatz
\be
\phi=\frac{1}{\sqrt{16\pi}}\phi(r)e^{-i \omega t}\,,
\ee
and expanding the equations of motion around $\phi_0=0$, we find
\beq
&&B'/B=(r/4)\left[C \omega^2\phi^2+(\phi')^2+B\left(4\beta\rho_{F}\phi^2+32\pi\rho_{F}\right)\right]\nn\\
&&+(1-B)/r\,,\label{ST_eq1}\\
&&C'/C=2/r+(Br)/2\left[2\beta\phi^2(\rho_{F}-P)+16\pi\rho_{F}-16\pi P\right]\nn\\
&&-2B/r\,,\\
&&\phi''=\beta B (\rho_{F}-3P)\phi-C\omega^2\phi-2\phi'/r+C'\phi'/(2C)\,,\\
&&2P'=-\left(P+\rho_{F}\right)\left(CB'-BC'\right)/(BC)\nn\\
&&-\beta\phi\phi'\left(P+\rho_{F}\right)/(16\pi)\,.\label{ST_eq2}
\eeq
Here, we consider the matter fields to be described by a perfect fluid. Note that in all the equations we are only considering terms up to order $\phi^2$. The method to find compact stars is the same as described in the previous Sections, so we will not dwell on it further. For the perfect fluid we will consider the polytropic equation of state $P=K\rho_F^{\gamma}$, with the parameters used in the previous Sections.

\begin{figure}[htb]
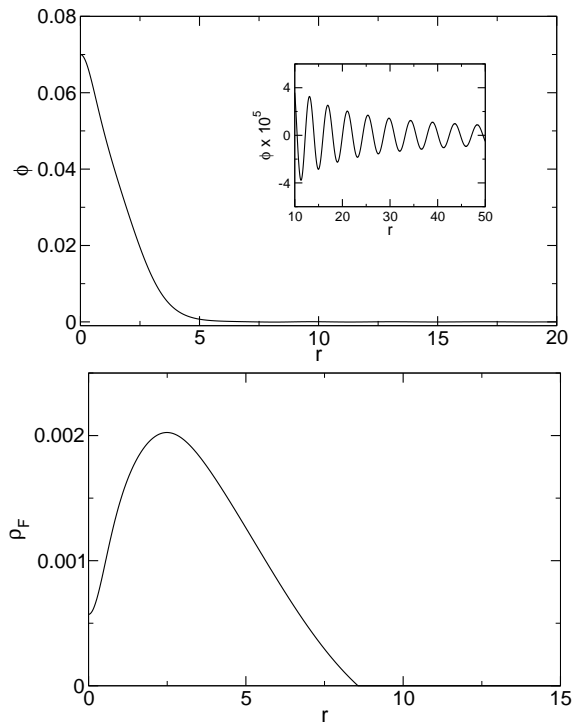

\begin{center}
\begin{tabular}{c}
\epsfig{file=scalartensor_phi,width=7.4cm,angle=0,clip=true}\\
\epsfig{file=scalartensor_rho,width=7.5cm,angle=0,clip=true}
\end{tabular}
\caption{Scalar field (top) and fluid's density (bottom) profiles for $\beta=7000$ and $\phi(0)=0.07$, by expanding the system~\eqref{ST_eq1}--~\eqref{ST_eq2} up to order $\phi^2$. For this solution we get $\omega=1.22$.
The inset of the left panel shows a zoom of the scalar field at large distances.\label{scalartensor_solution}}
\end{center}
\end{figure}
\begin{figure}[htb]
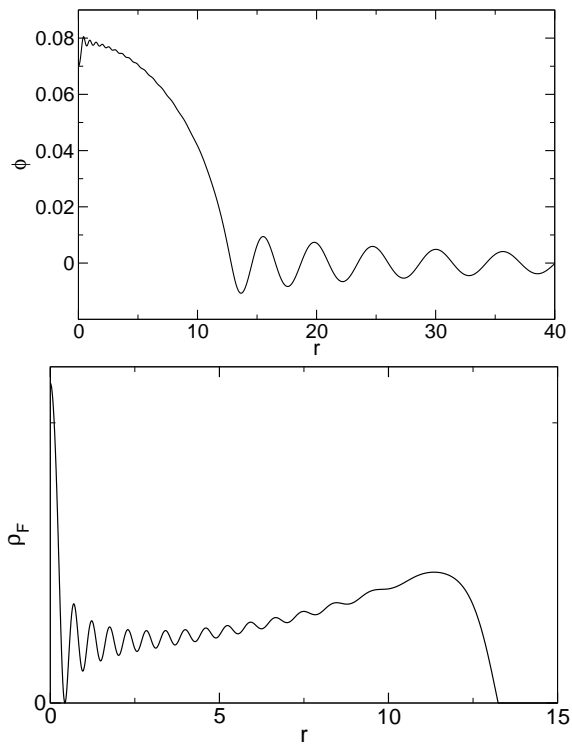

\begin{center}
\begin{tabular}{c}
\epsfig{file=scalartensor_exp_phi,width=7.4cm,angle=0,clip=true}\\
\epsfig{file=scalartensor_exp_rho,width=7.5cm,angle=0,clip=true}
\end{tabular}
\caption{Scalar field (top) and fluid's density (bottom) profiles for $A=e^{\beta|\phi|^2/2}$, $\beta=7000$ and $\phi(0)=0.07$. For some specific frequencies the scalar field acquires a non-trivial profile inside the star. However the amplitude of the radiating tail is non-negligible. 
\label{ST_total}}
\end{center}
\end{figure}
A solution is shown in Fig.~\ref{scalartensor_solution}. We have not been able to find solutions for which the scalar decays exponentially at infinity~\footnote{and thus truly stationary solutions. In other words, a time-varying scalar that decays as $1/r$ at large distances leads to a non-zero flux of energy at infinity}. However, we find that for some specific frequencies $\omega$ the scalar field is exponentially suppressed inside the star. Outside the star these solutions display an oscillating tail, indicating that they are not truly stable solutions but are instead long-lived solutions, slowly decaying through the emission of scalar radiation. This is very similar to what happens for oscillatons~\cite{Fodor:2009kg,Grandclement:2011wz}. We have only been able to find such solutions in the range $\beta\gg 1$. Negative values of $\beta$ are highly constrained by binary pulsar experiments~\cite{Damour:1996ke}, however positive values of $\beta$ remain unconstrained~\footnote{However see Refs.~\cite{Mendes:2014ufa,Palenzuela:2015ima} for a recent proposal to constrain positive values of $\beta$.}.

Although a careful analysis is out of the scope of this paper, our results make it possible that some massless scalar-tensor theories allow for the existence of stars with long-lived scalar cores. We would like to emphasize that our results are formally only valid up to order $\phi^2$. In the regime where such solutions exist, higher-order terms are in general important and should be taken into consideration. For the specific cases we tried, in particular $A=e^{\beta|\phi|^2/2}$, higher-order terms change drastically the solution as shown in Fig.~\ref{ST_total}. Although, for some specific frequencies, some solutions display a non-trivial profile inside the star, the amplitude of the radiative tail is non-negligible (and thus these solutions will dissipate over smaller time-scales). However full dynamical studies are needed to accurately compute the time-scale over which these configurations disperse.

\begin{figure*}[ht]
 \psfig{file=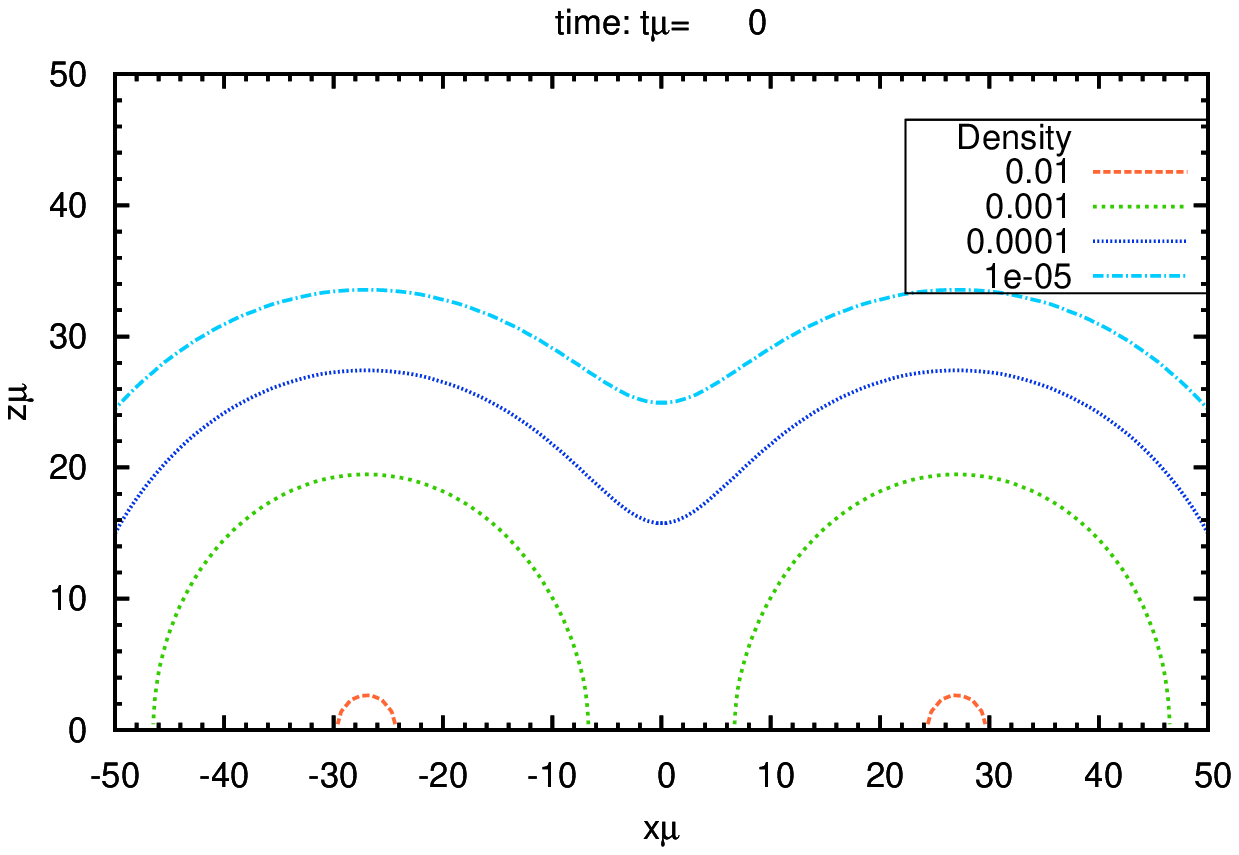,width=6.cm}
 \psfig{file=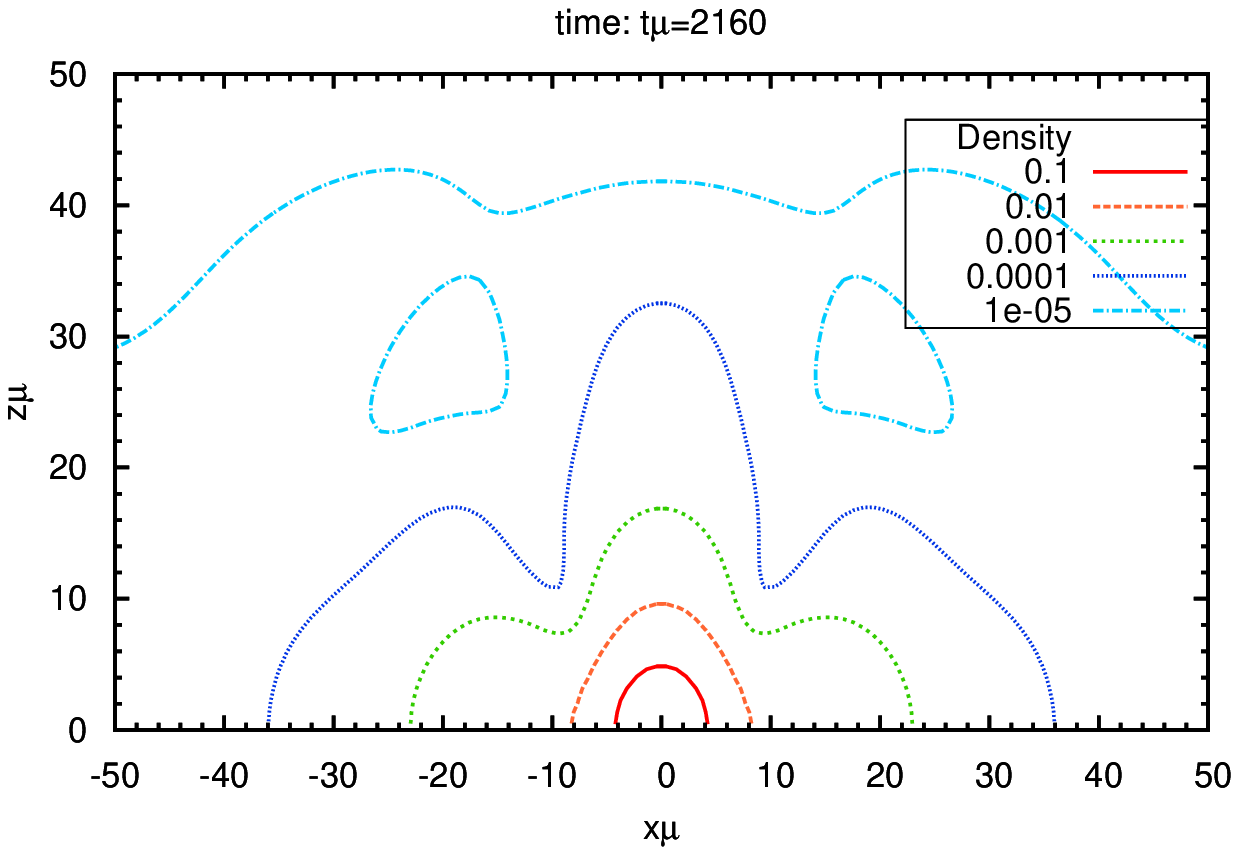,width=6.cm}
 \psfig{file=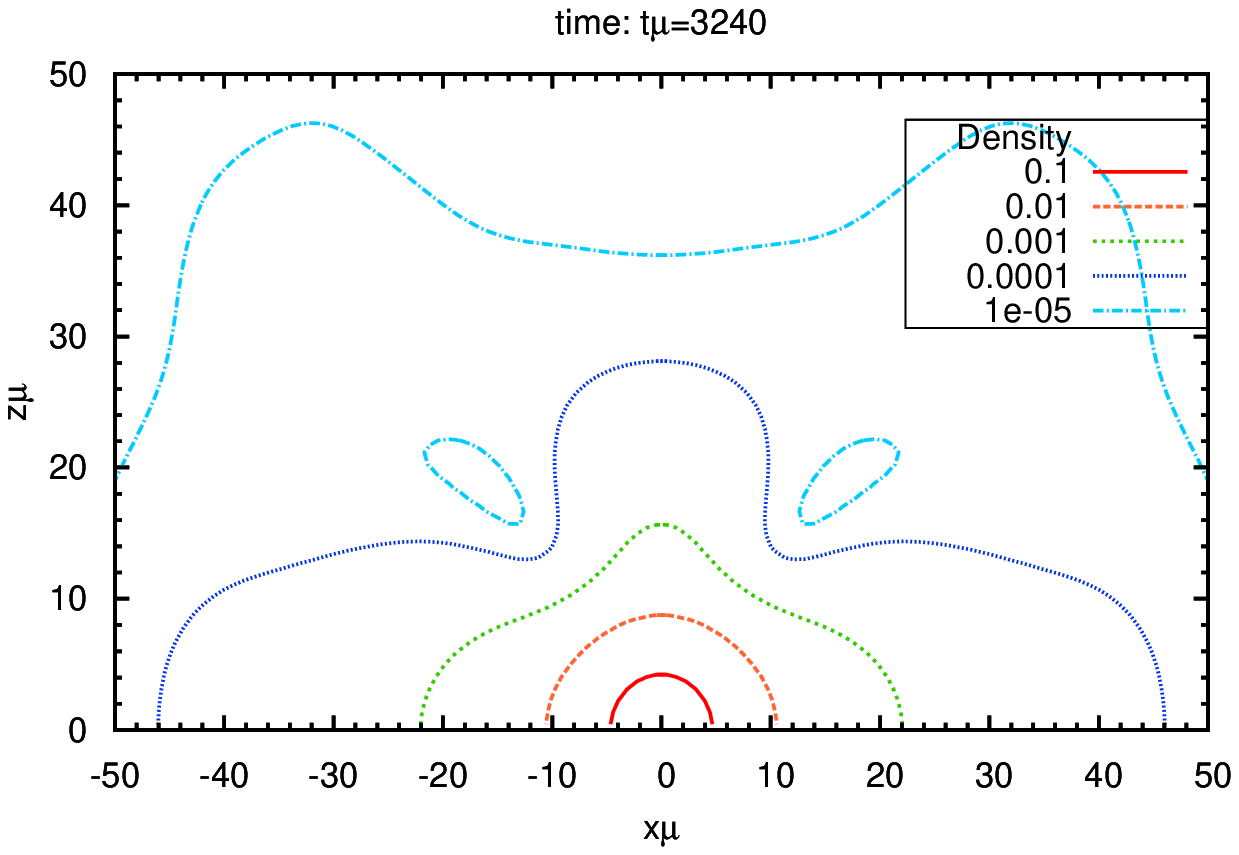,width=6.cm}\\
\psfig{file=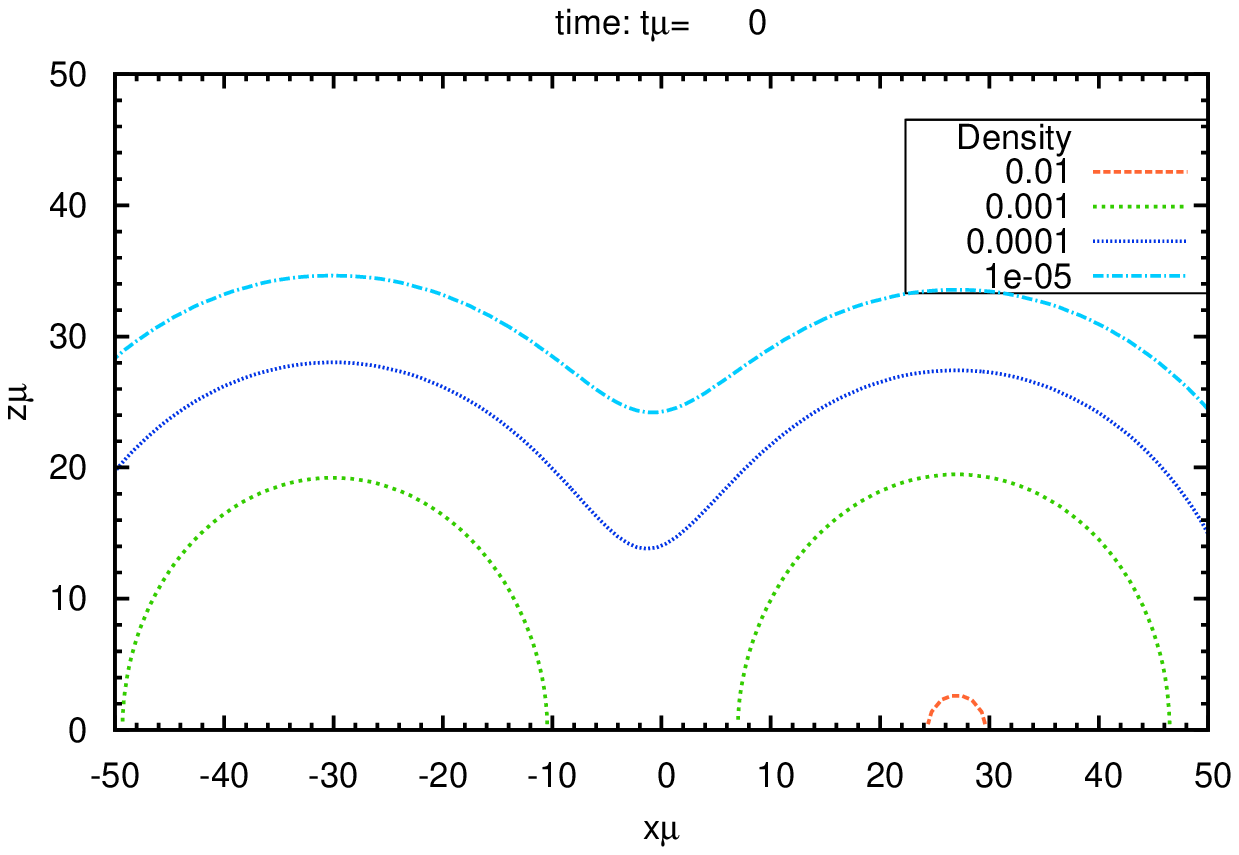,width=6.cm}
 \psfig{file=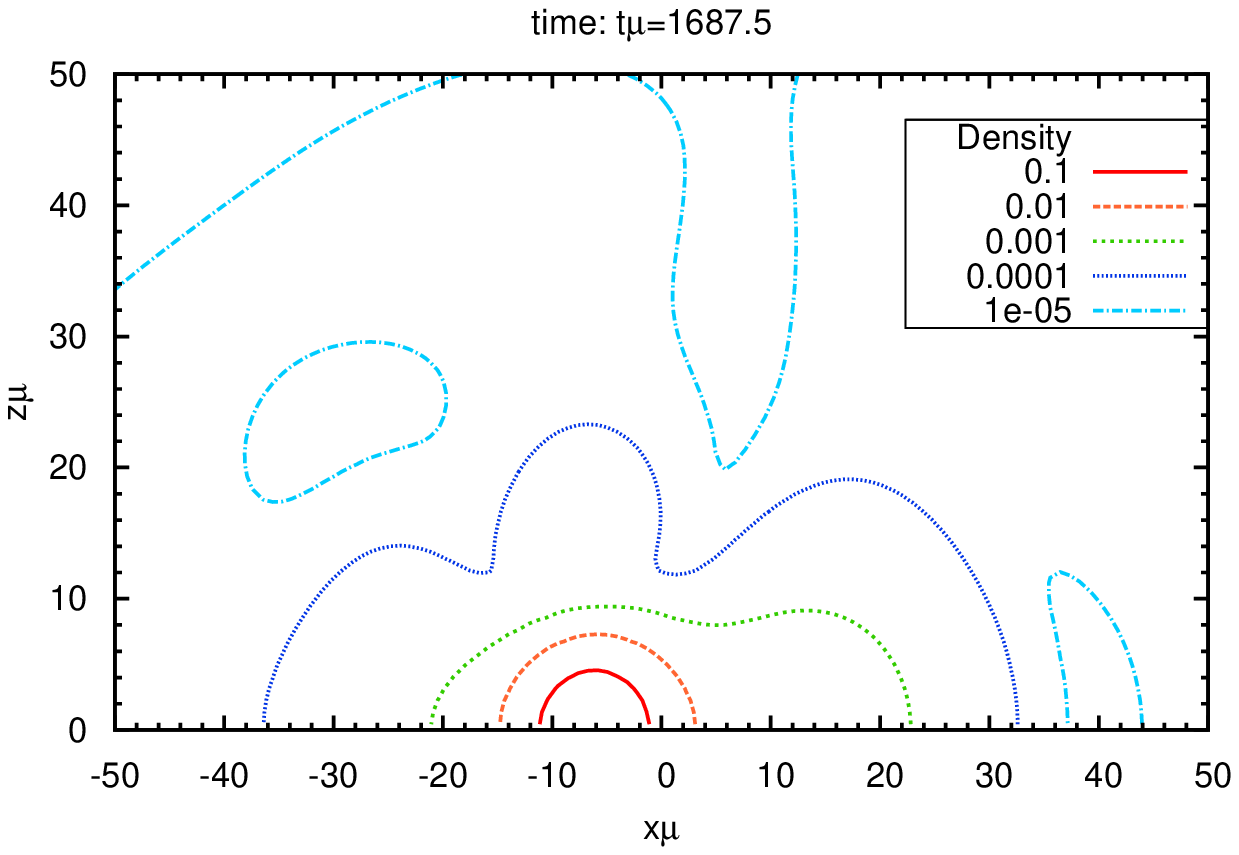,width=6.cm}
 \psfig{file=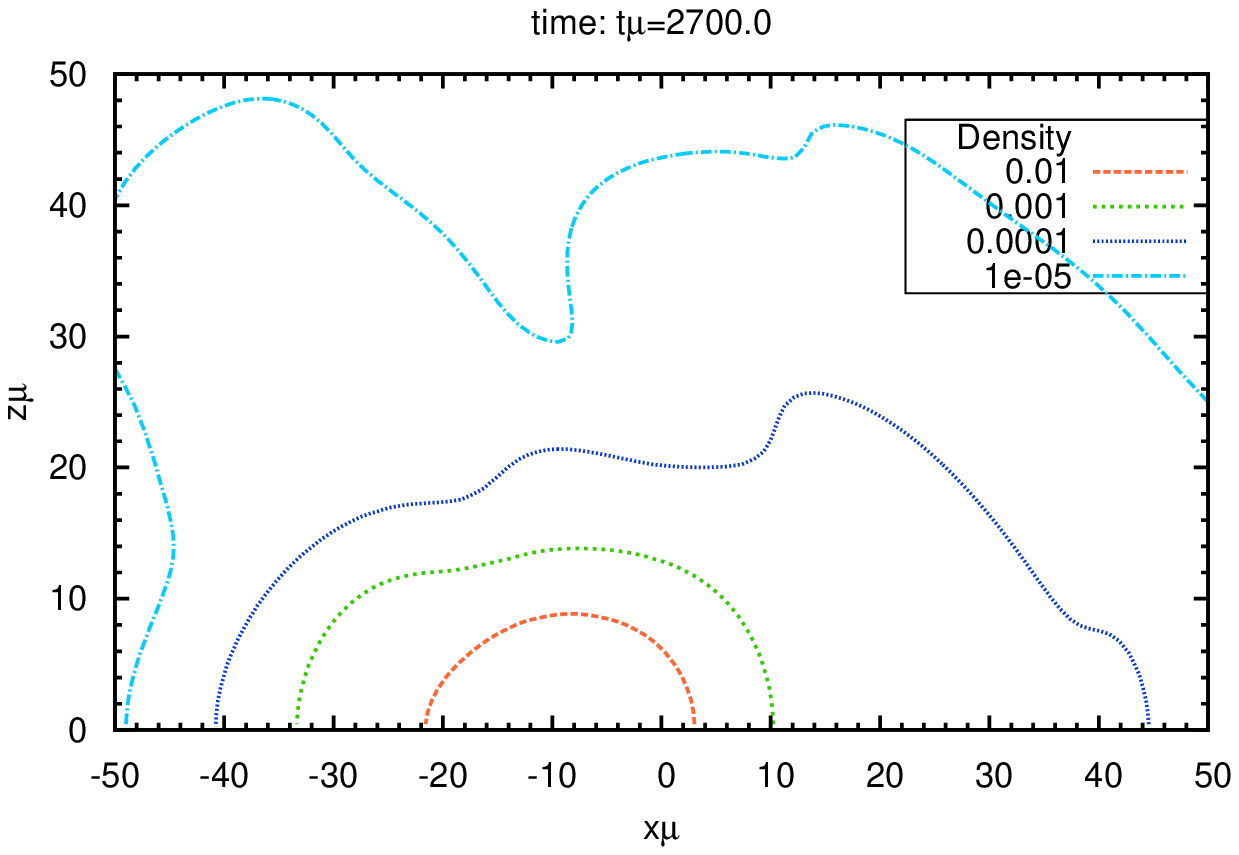,width=6.cm}\\
\psfig{file=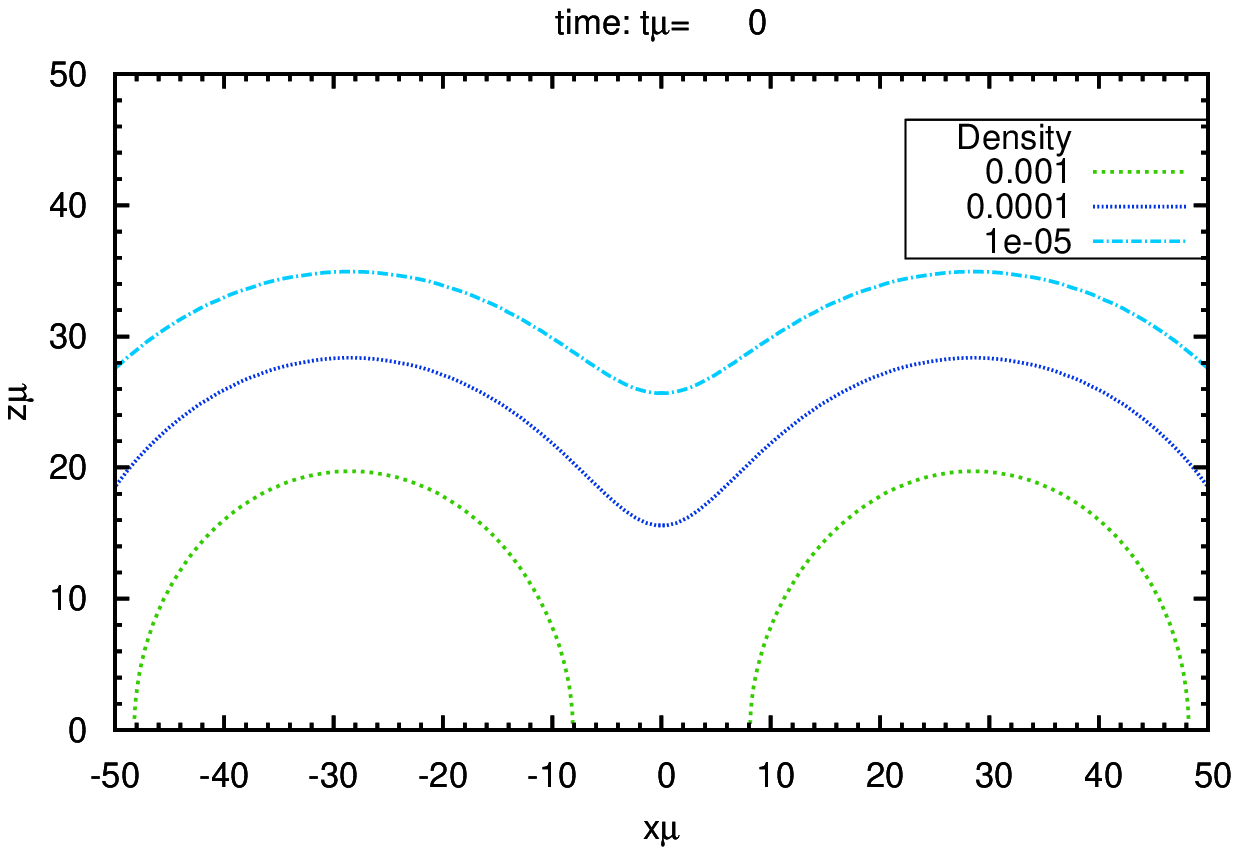,width=6.cm}
 \psfig{file=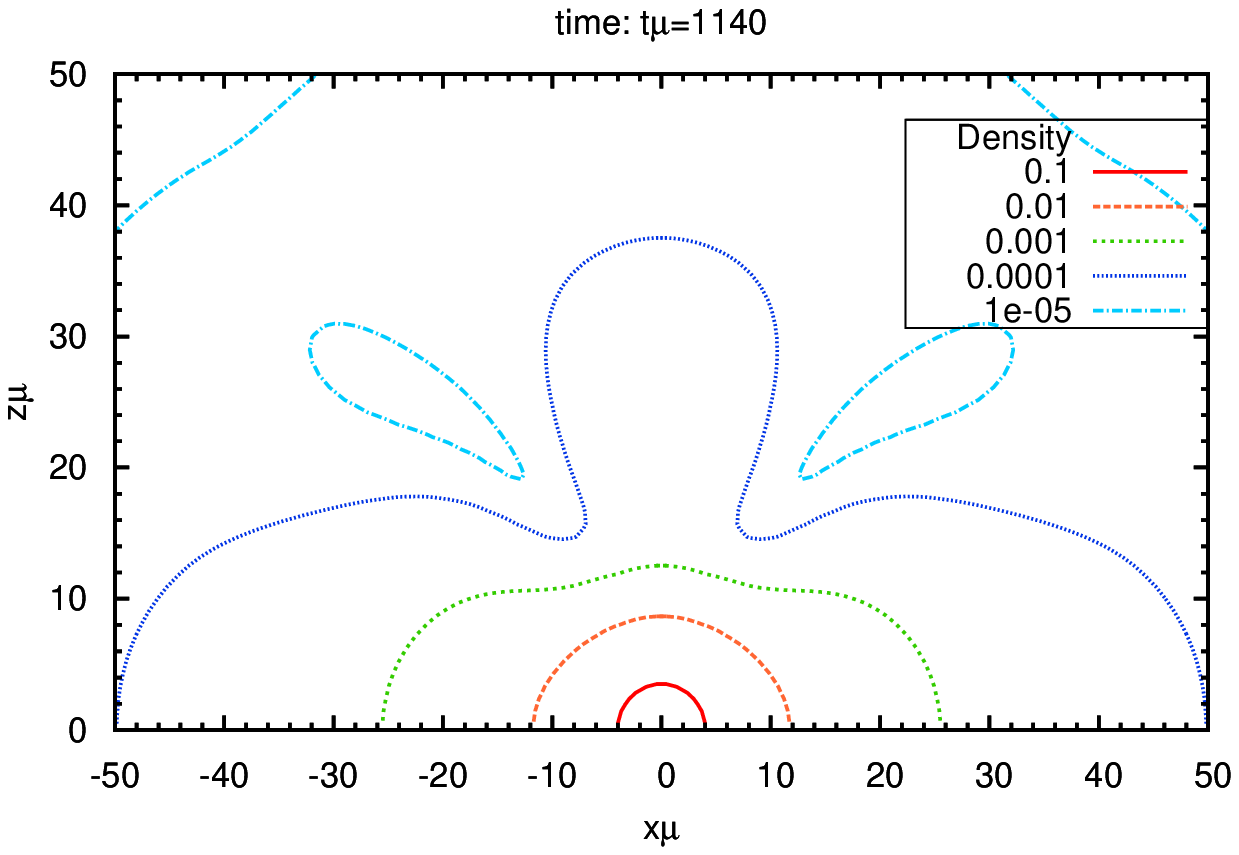,width=6.cm}
 \psfig{file=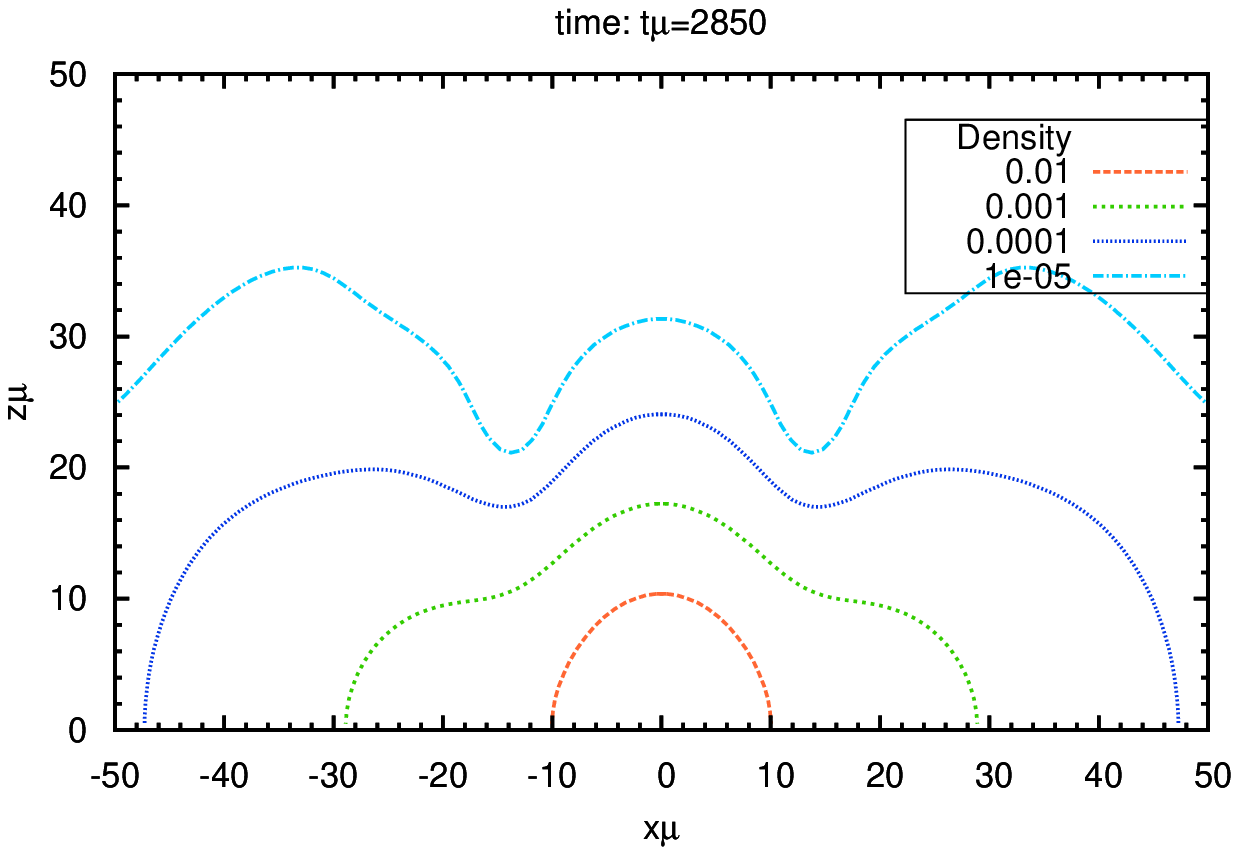,width=6.cm}\\
 \caption[]{Snapshots of the density $\rho_{\phi}$ during the collision of two oscillatons.
\noindent {\bf Upper panels:} equal-~mass oscillatons, each with $M_T\mu_S\sim 0.36, R\mu_S\sim 20$, along the plane of collision,
 for half the space (the remaining space can be obtained by symmetry). Note that the initial total mass of the system
is above the Chandrasekhar mass (i.e, the maximum mass in Fig.~\ref{MvsR}).
 At very late times, we find that it relaxes to a point on the curve of Fig.~\ref{MvsR}. 
\noindent {\bf Middle panels:} unequal-mass oscillatons
 with $M_T\mu_S\sim 0.36, R\mu_S\sim 20$ and $M_T\mu_S\sim 0.3, R\mu_S\sim 30$ along the plane of collision.
Again, the total mass is above the Chandrasekhar mass, but the final state is an oscillaton in the stable branch
of the diagram. 
\noindent {\bf Bottom panels:} equal-mass oscillatons
 each with $M_T\mu_S\sim 0.3, R\mu_S\sim 30$, along the plane of collision. Notice that the final star is {\it more massive} than any of the individual colliding stars, showing that it is possible to grow DM cores.
 }
 \label{fig:collisions}
\end{figure*}
%
\section{Accretion and growth of dark matter cores}\label{sec:collision}

We have shown that pulsating stars with DM cores exist as solutions of the field equations, even when the scalar DM core is self-gravitating. Although more studies are required, we also argued that these equilibrium solutions are stable. Do they form dynamically?
Pulsating purely bosonic states certainly do, through collapse of generic initial data~\cite{Seidel:1991zh,Garfinkle:2003jf,Okawa:2013jba}. There are two different channels
for formation of composite fluid/boson stars. One is through gravitational collapse in a bosonic environment, through which the star is born already with a DM core. The second process consists of capture and accretion of DM into the core of compact stars. A careful analysis for WIMPS has been done some time ago, showing that a significant amount of DM can be captured during the star's lifetime~\cite{Press:1985ug,Gould:1989gw}. The capture rate calculation for bosonic condensates follows through, {\it if} the condensate is small enough that it can be considered pointlike (we recall that bosonic condensates have a size determined by its total mass; very light condensates are spatially broad). If the condensate is ultralight and macroscopically-sized, interactions with the star are likely to be enhanced.
\subsection{Growth of DM cores for non-interacting fields}

Once the scalar is captured it will interact with the boson core. Interactions between complex fields have shown that
equal mass collisions at low energies form a bound configuration~\cite{Bernal:2006ci,Palenzuela:2006wp}. In other words, two bosonic cores composed of complex fields interact and form a more massive core at the center. This new bound configuration is in general asymmetric and will decay on large timescales~\cite{Macedo:2013jja}, the final state being spherical~\cite{Bernal:2006it}. 

The analysis of the previous Section makes it clear that for large and small boson masses, the boson and fluid behave as decoupled entities. As such, accretion by the DM core is well approximated by considering the collision between oscillatons. Using the methods of Numerical Relativity outlined in Refs.~\cite{Okawa:2014nda,Okawa:2014nea} (see also Appendix~\ref{app:num}), we have considered collisions between two scalar oscillatons for three different cases:

{\bf (i)} Two equal-mass oscillatons colliding at sufficiently small energies.
An example of such a collision is shown in the upper panels of
Fig.~\ref{fig:collisions} for oscillatons each with $M_T\mu_S\sim 0.36$
(corresponding to the curve $P_2$ in Fig.~\ref{fig:Single_Oscillaton}). 
Note that the total mass is larger than the peak value and one would naively predict gravitational collapse to a BH.
Instead, the final result is an oscillating object below the critical mass as shown by the red-solid curve in
Fig.~\ref{fig:collision_totalmass} (see also the convergence test in Appendix~\ref{app:num}).
\begin{figure}[htb]
 \epsfig{file=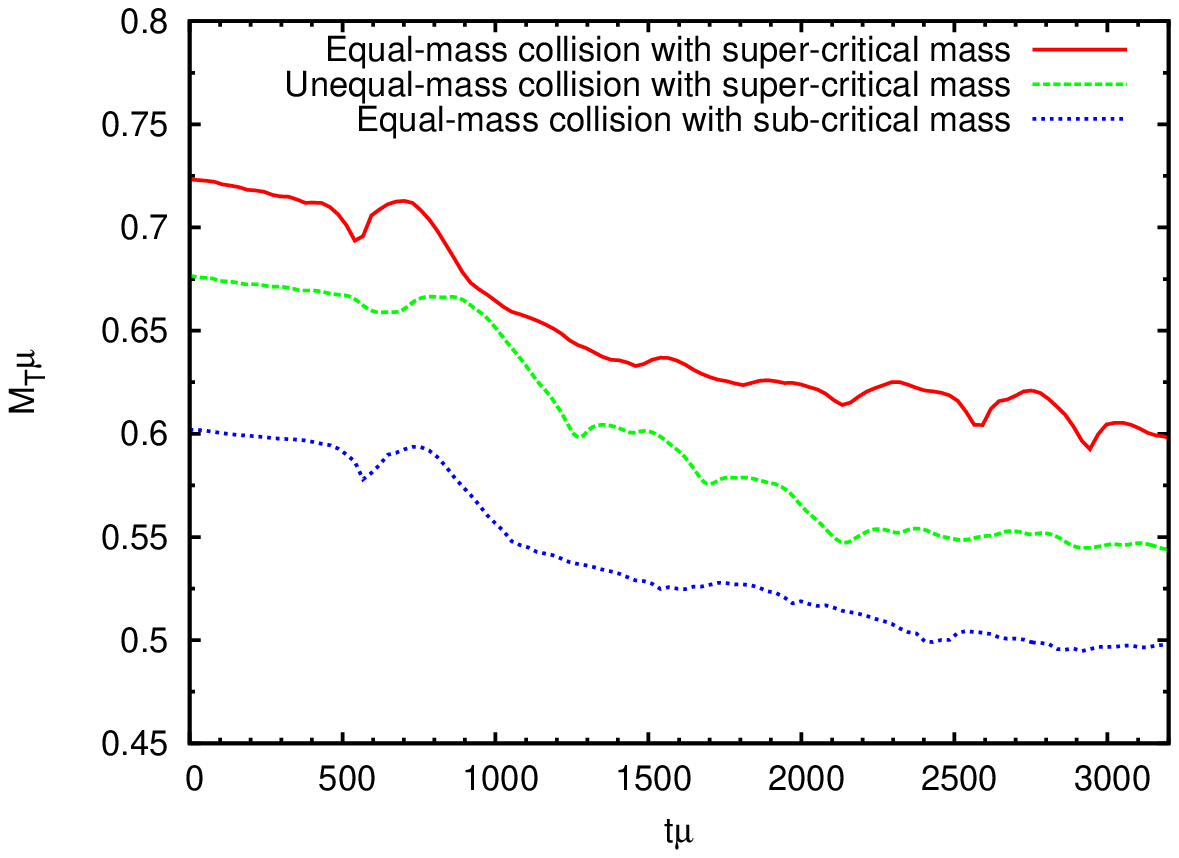,width=8.cm,angle=0,clip=true}
 \caption{Total mass of final object for the collision of scalar
 condensates, with total mass below and above the critical threshold
 $M_{\rm max}$. No BH formation is observed.}
 \label{fig:collision_totalmass}
\end{figure}

{\bf (ii)} Unequal-mass oscillatons for a total mass above the peak value.
The result of this collision is depicted in the middle panels of Fig.~\ref{fig:collisions}.
Also in this case the final configuration relaxes to a perturbed configuration
which oscillates around a stable configuration on the curve of Fig.~\ref{MvsR}.
The time evolution of the total mass is depicted by the green-dashed curve in Fig~\ref{fig:collision_totalmass},
where one can see that the total mass gradually decreases after the collision.

{\bf (iii)} Two equal-mass oscillatons but for a total mass below the peak value.
Even for this case, the final outcome is also an oscillating object below the critical mass
whose mass as a function of time is shown by the blue-dotted curve in Fig.~\ref{fig:collision_totalmass}.
It is also important to highlight that the final total mass is {\it higher} than those of the individual stars, showing that
cores {\it can} grow.

All these collisions share common features. In particular, one of our main results is that collapse to a BH is generically avoided for the cases where the total mass is larger than the critical stable mass \footnote{collapse to BHs {\it can} occur for certain special initial conditions, such as high-energy collisions of boson stars or even spherically symmetric collapse~\cite{Choptuik:2009ww,Palenzuela:2007dm,Okawa:2013jba}. These results are however not in contradiction with our statement and findings that, for accretion-related problems collapse to BHs seems to be avoided.}. This is a general feature of what has been termed the ``gravitational cooling mechanism'': a very efficient (dissipationless) mechanism that stops them from growing past the unstable point, through the ejection of mass~\cite{Seidel:1993zk,Alcubierre:2003sx,Guzman:2006yc}. Such features have been observed in the past in other setups, such as spherically symmetric gravitational collapse~\cite{Seidel:1993zk}
(see Fig. 2 in Ref.~\cite{Okawa:2013jba}), slightly perturbed oscillatons~\cite{Alcubierre:2003sx} or fields with a quartic self-interacting term~\cite{Guzman:2006yc}. Gravitational cooling provides a counter-example to an often-used {\it assumption} in the literature, that stars accreting DM will grow past the Chandrasekhar limit for the DM core and will collapse to a BH~\cite{Gould:1989gw,Bertone:2007ae,McDermott:2011jp,Graham:2015apa,Kouvaris:2013kra,Bramante:2014zca,Bramante:2015cua,Kouvaris:2011fi}. Our results show that this need not be the case, if the DM core is prevented from growing by a self-regulatory mechanism, such as gravitational cooling. In fact, avoidance of the BH final state has been seen in collisions of super-critical neutron stars as well~\cite{Kellermann:2010rt,Noble:2015anf}, which means that the phenomena is not exclusive of scalar fields.

\subsection{Effects of self-interaction}
%
\begin{figure}[ht]
 \psfig{file=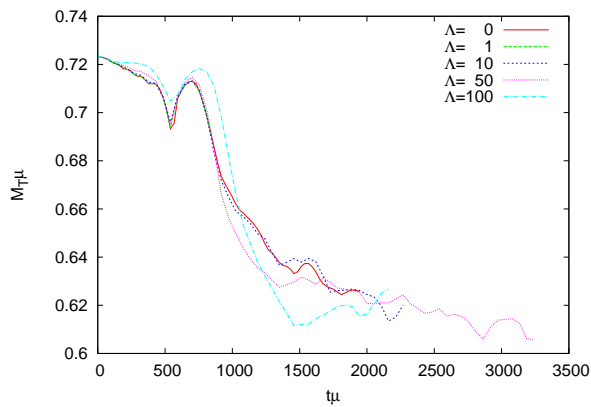,width=8cm}
 \caption[]{Time variation of total mass for the collision including self-interaction $\lambda$.
 Each initial oscillaton is described by the Gaussian $P_2$($Aw=0.9$ and $w\mu=18$) in Fig.~\ref{fig:Single_Oscillaton}.
 }
 \label{fig:Selfinteraction_Collision}
\end{figure}
The previous results concerned exclusively non-interacting fields (in the notation of Appendix~\ref{app:num}, $\lambda=0$).
We have extended our calculations to quartic self-interactions as well, the formalism is laid down in Appendix~\ref{app:num}.
In Fig.~\ref{fig:Selfinteraction_Collision}, we investigate the effects of self-interaction
on the evolution of the collision of two equal-mass oscillatons.
We define the parameter of self-interaction strength by $\Lambda\equiv\lambda/\mu^2$.
Our results show that the same qualitative features arise also for self-interacting fields, in particular strong gravitational-cooling effects.

Thus, even though other more detailed simulations are still needed, the likely scenario for evolution would comprise a core growth through minor mergers, slowing down close to the mass-radius peak (see Fig.~\ref{MvsR}), at which point it stops absorbing any extra bosons~\cite{Alcubierre:2003sx,Okawa:2013jba}. In other words, the unstable branch is never reached. This phenomenology is specially interesting, as it would also provide a capture mechanism for these fields which is independent of any putative nucleon-axion interaction cross-section: as we discussed, the bosonic core grows (in mass) through accretion until its peak value. At its maximum, it has a size $R_B/M_{\odot}\sim M_B/M_{\odot}$. This is the bosonic core {\it minimum size}, as described by Fig.~\ref{MvsR}. In other words, even for $M_B=0.01M_T$ the boson core has a non-negligible size and is able to capture and trap other low-energy oscillatons.
\subsection{Thermal effects inside stars}
Finally, we need to discuss thermal effects on the long-term evolution of these DM cores.
We have just learned that boson stars and oscillatons are prone to mass loss in highly dynamical situations.
It is, in principle, possible that thermal motion of the star's material alone is enough to disperse completely
a scalar condensate.

\begin{figure}[ht]
 \psfig{file=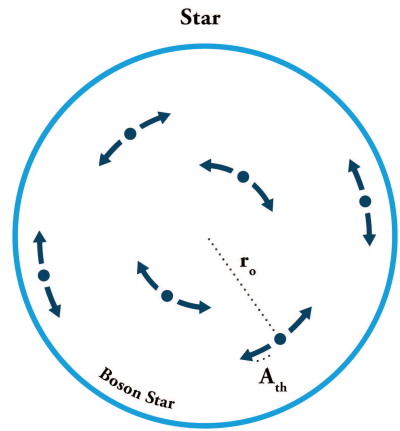,width=6.cm}
 \caption[]{Cartoon depicting the thermal motion of nucleons inside stars, a fraction of which may be occupied by a boson star.
 The motion of the nucleons is modelled as harmonic and is due to collisions with other nucleons.
 }
 \label{fig:thermal}
\end{figure}
We will investigate this issue in a conservative setup, where we turn off all non-gravitational
interactions between the oscillaton and the host star. Focus first on a single nucleon in the star undergoing thermal
agitation of amplitude $A_{\rm th}$, see Fig.~\ref{fig:thermal}. The density of the star material can be related to $A_{\rm th}$ and each nucleon's mass $m_N$ via $\rho=m_N/A_{\rm th}^3$. For simplicity, we model the nucleon's position to be given by a fixed radius $R$ and an azimuthal angle
\be
\varphi=\frac{A_{\rm th}}{R}\sin{\Omega_{\rm th} t}\,. \label{motion_thermal}
\ee
The frequency $\Omega_{\rm th}$ of the nucleon can be related to its mean free path $A_{\rm th}$ via $m_N(A_{\rm th}\Omega_{\rm th})^2=k_B T$.
It is clear that emission of gravitational and scalar waves is suppressed for energy scales $\Omega \ll \mu_S$, since the excitations are then confined to the interior of the boson star or oscillaton. However, for $\Omega_{\rm th} \gg \mu_S$ the fluctuations induced by the thermal motion are free to propagate to infinity and leave the star, leading to mass loss of the oscillaton. In other words,
for very light fields, $\mu_S \lesssim 10^{-5} \, {\rm eV}$ and realistic star temperatures $T\gtrsim 10^5\,{\rm K}$, a nonzero mass loss occurs, which we now need to estimate.
Notice that we neglect the tail of the Maxwell-Boltzmann distribution, and therefore our estimate will be doubly conservative.

Our approach will give us an order of magnitude estimate of the scalar radiation loss due to the thermal motion of the star material. We will make two assumptions:

\noindent {\bf i.} The stresses causing the nucleon to move as in \eqref{motion_thermal} can be neglected. Effectively then, to compute the radiation released in the process,
one only uses the stress-energy tensor of the nucleon. This approximation neglects whatever is causing the (non-geodesic) motion. This assumption will cause factors of order unity
to be neglected~\cite{Price:1973ns}. It does not, in principle, affect the order of magnitude of the numbers we show.

\noindent {\bf ii.} The thermal motion of {\it all} the nucleons is taken into account by adding incoherently the radiation from a single nucleon.
This is, in principle, a reasonable assumption and is taken on the grounds that the motion of the nucleons is random.
This assumption implies that if one nucleon radiates a scalar flux $\dot{E}_1$ then the total scalar flux is $\dot{E}_N=N\dot{E}_1$, with $N$ the total number of nucleons effectively contributing to the flux.

The details of our calculations are given in Appendix~\ref{app:thermal}. We find that the flux of scalar radiation at infinity scales as
\be
\dot{E}_{1}\sim m_N^2\left(\frac{A_{\rm th}}{r_0}\right)^2\left(\Omega_{\rm th}/\mu_S\right)^{4}\,,
\ee
for large $\Omega_{\rm th}/\mu_S$. The lifetime of a condensate with mass $M_B$, due to these thermal effects, can be estimated as $\tau \sim M_B/\dot{E}$; even for ultralight fields and compact and large condensates this
timescale is much larger than the age of the universe. 

In summary, in this setup where non-gravitational interactions are shut-off, thermal effects have no impact on the dynamics of DM condensates at the center of stars.
\section{Conclusions}\label{sec:conclusion}

The main purpose of this work (and of the recent \emph{Letter}~\cite{Brito:2015yga}) is to understand how DM might affect the structure of compact stars, using a fully relativistic setup. 
The picture we discussed is quite generic and shows that any self-gravitating, massive bosonic field can form compact structures, either in the form of boson stars or oscillatons.
In particular, we showed for the first time that real self-gravitating massive vector fields can form oscillatons and boson stars~\cite{Brito:2015pxa}.

These structures can cluster inside stars leading to oscillating configurations with distinctive imprints. 
In particular, since the fundamental frequency is $\omega\sim \mu_{S,\,V}$ for non-compact stars (cf. Section~\ref{sec:fluid}), these oscillations imply that both the bosonic field and the fluid density (which is coupled to it gravitationally), vary periodically with a frequency
\be
f=2.5\times 10^{14}\,\left(\frac{m_{B}c^2}{eV}\right)\,{\rm Hz}\,,
\ee
or multiples thereof. For axion-like particles with masses $\sim 10^{-5} \,eV/c^2$, these stars would emit in the microwave band.
These oscillations are driven by the boson core and might have observable consequences; it is in principle even possible that resonances occur when the frequency of the scalar is equal to the oscillation frequency of the unperturbed star. 
The joint oscillation of the fluid and the boson might be called a {\it global thermalization of the star}, and is expected to occur also
for boson-star-like cores (which give rise to static boson cores), once the scalar is allowed to have non-zero couplings with the star material. Such couplings where recently considered
in Ref.\cite{Arvanitaki:2015iga}, who showed that the oscillations could leave imprints in the Earth's breathing modes and possibly be observable with Earth-based detectors. Although further work is needed, our analysis shows that stars could also work as good DM detectors. Additional signatures could also occur in the presence of a DM core. For example, bosonic cores are intrinsically anisotropic. A certain degree of anisotropy in a neutron star might leave important imprints that could potentially be measured~\cite{Silva:2014fca} (some effective field theories going beyond the mean-field approximation also predict that inside neutron stars, nuclear matter may become anisotropic, see e.g~\cite{Adam:2010fg,Nelmes:2012uf,Adam:2014dqa,Adam:2015lpa}. The interplay between these two effects is also an interesting subject for future research).

Using full nonlinear evolutions of the field equations and arguments based on the structure of these DM cores, we showed that composite structures can be stable, even when the DM core is self-gravitating.
We also provide clear and precise criteria for the onset of instabilities.

We should stress that previous works on the subject of DM accretion by stars have implicitly assumed that
the DM core is able to grow without bound and eventually collapse to BHs~\cite{Gould:1989gw,Bertone:2007ae,McDermott:2011jp,Graham:2015apa,Kouvaris:2013kra}.
Our results, from full nonlinear simulations of the field equations, show that the core may stop growing when it reaches a peak value, at the threshold
of stability, if DM is composed of light massive fields. Gravitational cooling quenches the core growth for massive cores and the core growth halts, close to the peak value (c.f. Fig.~\ref{MvsR}). Similar mechanisms have also been shown to  be effective in collisions of super-critical neutron stars~\cite{Kellermann:2010rt} (see also~\cite{Noble:2015anf} and references therein), and thus this phenomena is a very generic feature of self-gravitating solutions, and not only of bosonic fields. Together with our work, these studies show that BH formation depends very sensitively on the initial conditions of the system and cannot be solely inferred from the linear stability of the stars.

Our study focused on core growth through lump accretion, and does not address other forms of growth, in particular 
more continuous processes like spherical accretion or wind accretion. Partial results in the literature indicate that gravitational cooling mechanisms are also active in these setups~\cite{Hawley:2000dt}.
To study BH formation in these different scenarios, a complete scan of the parameter space would be necessary. This could be done along the lines of Refs.~\cite{Hawley:2000dt,Noble:2015anf}.

Future works should consider more realistic equations of state and possibly include viscosity in the star's fluid and local thermalization. Viscous timescales for neutron star oscillations can be shown to be large compared to the star dynamical timescale $R$,
but small when compared to the (inverse of) the accretion rate likely to be found in any realistic configuration~\cite{1990ApJ...363..603C}. As such, we expect that viscosity will damp global oscillations of the star, eventually leading to a depletion of the scalar field core. A similar effect will occur with local thermalization of the scalar with the star material if the central temperature of the star is much larger than the mass of the bosonic field~\cite{Bilic:2000ef,Latifah:2014ima}. On the other hand, although more detailed studies of these effects are still necessary, the results of Refs.~\cite{Bilic:2000ef,Latifah:2014ima} suggest that, for bosonic fields with masses $\gtrsim$ keV inside old neutron stars or white dwarfs, local thermalization should not significantly affect our results.

We also argued that in theories where a scalar field acquires an effective mass due to the presence of matter, long-lived oscillating configurations might form inside stars, in a region of the parameter space which remains unconstrained. Whether these configurations actually form and whether they have peculiar observable imprints, can only be accessed through a fully dynamical analysis. On the other hand, our results apply directly to massive scalar-tensor theories (see e.g. Ref.~\cite{Chen:2015zmx}). Further work is still needed, but our results raise the interesting possibility that DM cores could be used to further constrain these theories.

There are a number of other setups where similar results may hold. For example, minimally coupled, multiple (real) scalars, interacting only gravitationally, were also shown to give rise to similar configurations~\cite{Hawley:2002zn}. In higher dimensions, one may ask if purely gravitational oscillatons exist. Such solutions could arise due to the compactification of extra dimensions, which effectively give rise to massive bosonic fields. A natural extension of our work would be to look for such solutions in theories with massive gravitons~\cite{deRham:2014zqa}, which have received considerable attention in the past few years.
Another outstanding open problem concerns the construction of rotating oscillatons. Rotating boson stars were obtained for both complex scalar fields~\cite{Yoshida:1997qf} and more recently for complex vector fields~\cite{Brito:2015pxa}. For real fields, the time-dependence of the metric makes the explicit construction of rotating oscillatons an highly intricate task. A possibility would be to use Numerical Relativity methods to construct such solutions. We hope to possibly solve some of these problems and further develop this subject in the near-future. 

\begin{acknowledgments}
We thank Daniela Alic, Emanuele Berti, Joseph Bramante, Leonardo Gualtieri, Ilidio Lopes, David Marsh, Shinji Mukohyama, Frans Pretorius and Ulrich Sperhake for useful comments and feedback.
R.B. acknowledges financial support from the FCT-IDPASC program through the grant SFRH/BD/52047/2012, and from the Funda\c c\~ao Calouste Gulbenkian through the Programa Gulbenkian de Est\' imulo \`a Investiga\c c\~ao Cient\'ifica.
V.C. thanks the Departament de F\'{\i}sica Fonamental at Universitat de Barcelona for hospitality while this work was being completed.
V.C. acknowledges financial support provided under the European Union's H2020 ERC Consolidator Grant ``Matter and strong-field gravity: New frontiers in Einstein's theory'' grant agreement no. MaGRaTh--646597. Research at Perimeter Institute is supported by the Government of Canada through Industry Canada and by the Province of Ontario through the Ministry of Economic Development $\&$
Innovation.
V.C. also acknowledges financial support from FCT under Sabbatical Fellowship nr. SFRH/BSAB/105955/2014.
CP acknowledges support from the Spanish Ministry of Education and Science through a Ramon y Cajal grant and from the Spanish Ministry of Economy and Competitiveness grant FPA2013-41042-P. 
C.M.  thanks the support from Conselho Nacional de Desenvolvimento Cient\'ifico e Tecnol\'ogico (CNPq).
This work was supported by the H2020-MSCA-RISE-2015 Grant No. StronGrHEP-690904.
The authors thankfully acknowledge the computer resources, technical expertise and assistance provided by CENTRA/IST. Computations were performed at the clusters
``Baltasar-Sete-S\'ois'' and Marenostrum, and supported by the MaGRaTh--646597 ERC Consolidator Grant.
\end{acknowledgments}
\clearpage
\newpage

\appendix
\section{Oscillating nucleons and DM loss\label{app:thermal}}
%
\begin{figure}[ht]
 \psfig{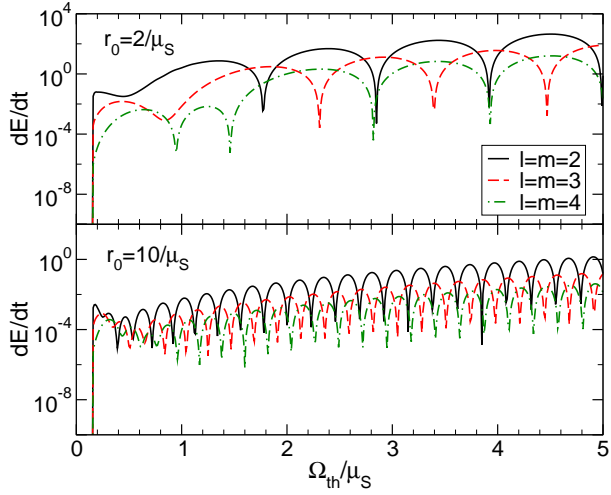}
 \caption[]{Energy flux from a (fluid) point particle vibrating inside a boson star of radius $R\mu_S=7.84$, and at a distance $r_0$ away from the center. The flux is shown as a function of the point particle's vibration frequency $\Omega_{\rm th}$. At large frequencies the flux scales
as $\dot{E}\sim \Omega_{\rm th}^{4}$.
 }
 \label{fig:scalar_rad_thermal}
\end{figure}
In order to investigate the mass loss via thermal effects, we consider a single nucleon vibrating, due to thermal effects, with a small amplitude $A_{\rm th}$ inside the DM core. This nucleon is at a fixed radius $r_0$, but its azimuthal angle changes as
\be
\varphi = \frac{A_{\rm th}}{r_0}\sin (\Omega_{\rm th} t).
\ee
The formalism to study the perturbations induced by particles at a fixed radius in a bosonic condensate was developed in Refs.~\cite{Macedo:2013jja,Macedo:2013qea}. Since we are interested in the DM mass loss, we focus only in the polar sector of the perturbations. 
We shall assume the metric to be the one for a mini boson star, with maximum mass configurations, described by the line element
\be
ds^2=-F(r)dt^2+B(r) dr^2+r^2d\Omega^2.
\ee
Generically, the relevant coefficients of the expansion in the tensorial spherical harmonics (in the time domain) for a particle moving inside the star configuration are given by
\begin{align}
A_{lm}^{(1)}&=i\frac{\sqrt{2}}{r^2}m_N\gamma \sqrt{FB}\frac{dr}{dt} \delta(r-r_p)Y^{lm*}_p,\\
B_{lm}&=m_N \gamma  \sqrt{\frac{2}{l(l+1)}}r^{-1}\frac{dr}{dt}\sqrt{\frac{B}{F}} \delta(r-r_p)\frac{d}{dt}Y^{lm*}_p,\\
B_{lm}^{(0)}&=im_N \gamma  \sqrt{\frac{2}{l(l+1)}}r^{-1}\sqrt{\frac{F}{B}} \delta(r-r_p)\frac{d}{dt}Y^{lm*}_p,\\
F_{lm}&=m_N \gamma \sqrt{\frac{2(l-2)!}{(l+2)!FB}} \delta(r-r_p)\times \nonumber\\
&\l(\frac{1}{2}\l[{(\frac{d\theta}{dt}})^2-({\frac{d\varphi}{dt}})^2\sin^2\theta_p\r]{W_{lm}}^*_p+ \frac{d\theta}{dt}\,\frac{d\varphi}{dt}\,{X_{lm}}^*_p\r)\,,
\end{align}
where the subscript $p$ stands for the nucleon quantities, $Y^{lm}_p\equiv Y^{lm}(\theta_p,\varphi_p)$ are the spherical harmonics, and $\gamma=dt/d\tau$. The angular quantities $W_{lm}$ and $X_{lm}$ are defined in Refs.~\cite{Martel:2003jj,Sago:2002fe}. To specialize to our model for a fixed radius $r_p=r_0$, we can, without loss of generality, take the motion to be in the plane  $\theta_p=\pi/2$. Additionally, we have
\begin{align}
	\frac{dr}{dt}&=0,~\frac{d\varphi}{dt}=\frac{A_{\rm th}}{r_0}\Omega_{\rm th}\cos(\Omega_{\rm th} t),~\frac{d\theta}{dt}=0,\\
	\gamma&=\left[F(r_0)-r_0^2\frac{d\varphi}{dt}^2\right]^{-1/2}.
\end{align}
Note that the nucleon frequency is restricted to $A_{\rm th}\Omega_{\rm th}<F(r_0)$. Therefore, for an oscillating nucleon at a fixed radius, we have
\begin{align}
	A_{lm}^{(1)}&=0, ~ B_{lm}=0,\\
B_{lm}^{(0)}&=im_N m \gamma  \sqrt{\frac{2F}{l(l+1)B}}r^{-1}\delta(r-r_0)\nonumber\\
&e^{-i m\varphi}\frac{d\varphi}{dt}Y^{l0*}(\pi/2),\\
F_{lm}&=m_N \gamma \sqrt{\frac{(l-2)!}{2(l+2)!FB}}[l(l+1)-2m^2]\nonumber\\
&\delta(r-r_0)\left(\frac{d\varphi}{dt}\right)^2e^{-i m \varphi}Y^{l0*}(\pi/2).
\end{align}
To extract the corresponding source terms in the frequency domain, one has to perform the integral $\int dt (\star) e^{i \omega t}$ (cf. Refs.~\cite{Macedo:2013jja,Macedo:2013qea}). For a particle in a circular motion this is a direct task, because the time dependence is on $\varphi$, as $e^{-i m \Omega_{\rm th} t}$, and the integral results in a delta function $\delta(\omega-m\Omega_{\rm th})$. For the oscillatory motion, the time dependence is still encoded into the azimuthal angle, but the integration is not direct. In order to perform the integrations, we can use
\begin{align}
&e^{-i m \varphi}=\sum_{n=0}^\infty\sum_{k=0}^n{\cal A}_{nk}e^{-i(2k-n)\Omega_{\rm th} t},\\
\end{align}
where
\begin{align}
	{\cal A}_{nk}= \frac{(-1)^{k+n}}{2^nk!(n-k)!}\left(\frac{m A_{\rm th}}{r_0}\right)^n
\end{align}
together with the redefitions of the trigonometric functions in terms of exponentials, expanding in powers of $A_{\rm th}/r_0$. At first order in $A_{\rm th}/r_0$, the only non vanishing source function is
\begin{align}
	B_{lm}^{(0)}&={\cal F}(r,r_0)(e^{i\Omega_{\rm th} t}+e^{-i\Omega_{\rm th} t}) \delta(r-r_0),
\end{align}
where
\be
{\cal F}(r,r_0)=-\frac{A_{\rm th}}{r_0}m_N\frac{\gamma m \Omega_{\rm th}}{r\sqrt{2l(l+1)}}\sqrt{\frac{F}{B}}Y^{l0*}(\pi/2),
\ee
where at this order $\gamma=F(r_0)^{-1/2}$. Therefore, at first order, the time integral (and the source term in the frequency domain) results in two terms proportional to $\delta(\omega\pm\Omega_{\rm th})$. To integrate the differential equations, we can use the same method exploited in Refs.~\cite{Macedo:2013jja,Macedo:2013qea}. The energy flux is given by
\be
\frac{dE}{dt}=(\Omega_{\rm th}+\omega_0)k_+|\phi_+(r\to\infty)|^2
\ee
where $\omega_0$ is the frequency of the background scalar field, $k_+=\sqrt{(\Omega_{\rm th}+\omega_0)^2-\mu_S^2}$, and $\phi_+$ describes the perturbation induced by the moving nucleon (radiating DM field). From the above expression it is clear that the scalar radiation is suppressed for $\Omega_{\rm th}\ll \mu_S$. Scalar radiation appears for values of $\Omega_{\rm th}$ such that $(\Omega_{\rm th}+\omega_0)^2>\mu_S^2$.
Our numerical results are summarized in Fig.~\ref{fig:scalar_rad_thermal}, and are consistent with a dependence $\dot{E}\sim m_N^2A_{\rm th}^2/r_0^2 \left(\Omega_{\rm th}/\mu_S\right)^4$ at large frequencies.

\section{Head-on collision of oscillatons}\label{app:num}
%
In this section, we review the time evolution scheme using numerical relativity, for a real scalar field minimally coupled to Einstein's equations.
For completeness, we also consider self-interactions of the form $\lambda \phi^4$, as discussed in the main text.
In particular, we focus on the theory
\begin{eqnarray}
 S = \int d^4x
  \left( \frac{R}{2\kappa} -\frac{1}{2}\nabla^{\mu}\phi\nabla_{\nu}\phi
   -\frac{1}{2}\mu_S^2\phi^2 -\frac{1}{4}\lambda\phi^4\right)\,.\nonumber
\end{eqnarray}
%
\subsection{Time evolution formulation}
Let us first consider a generic line element in four-dimensional spacetimes given by
\begin{eqnarray}
 \dif s^2 &=& -\left(\alpha^2-\gamma_{ij}\beta^i\beta^j\right)\dif t^2
  +2\gamma_{ij}\beta^j\dif x^i\dif t +\gamma_{ij}\dif x^i\dif x^j,\nonumber\\\label{eq:line_element}
\end{eqnarray}
where $\alpha, \beta^i$ and $\gamma_{ij}$ denote the lapse function, the shift vector and induced metric, respectively.
We introduce the conjugate momenta for the metric and a massive scalar field~$\phi$ minimally coupled to gravity ,
\begin{eqnarray}
 K_{ij} = -\frac{1}{2\alpha}\ldif\gamma_{ij},\quad
 \Pi = -\frac{1}{\alpha}\ldif\phi,
\end{eqnarray}
where $\ldif$ is the Lie derivative along the shift vector given by $\ldif\equiv\partial_t-\Lie_{\beta}$.
In the ADM formalism, we describe the time evolution for these conjugate momenta using 
\begin{eqnarray}
 \ldif K_{ij} &=& - D_{i} D_{j} \alpha + \alpha\left( R_{ij} -2 K^{k}{}_{i} K_{jk} + K K_{ij} \right)\nonumber\\
 && + 4\pi\alpha\left(\gamma_{ij}(S-\rho) - 2S_{ij} \right),\label{eq:EvolKADM}\\
 \ldif \Pi &=& 
 \alpha\left( -D^{i}D_{i}\phi + K \Pi + \mu_S^2\phi+\lambda\phi^3\right)
 -D^{k}\alpha D_{k}\phi,\label{eq:EvolKphiADM}\nonumber
\end{eqnarray}
where $R_{ij}$ is the three-dimensional Ricci tensor and projected quantities of the energy momentum tensor are given by
\begin{eqnarray}
 \rho &=& \frac{1}{2}\Pi^{2} +\frac{1}{2}D^k\phi D_k\phi +\frac{1}{2}\mu_S^2\phi^2+\frac{1}{4}\lambda\phi^4,\label{eq:def_rho}\nonumber\\
 j_i &=& \Pi D_{i}\phi,\label{eq:def_j}\nonumber\\
 S_{ij} &=& D_i\phi D_j\phi +\frac{1}{2}\gamma_{ij}\left(\Pi^{2} -D^k\phi D_k\phi -\mu_S^2\phi^2-\frac{1}{2}\lambda\phi^4\right)\nonumber
  .\label{eq:def_sij}
\end{eqnarray}
The remaining equations by the 3+1 decomposition yield the Hamiltonian and momentum constraints,
\begin{eqnarray}
 \H &\equiv& R + K^2 - K_{ij} K^{ij} - 16\pi \rho = 0,\label{eq:constr_H}\nonumber\\
 \M_{i} &\equiv& D_{j} K^{j}{}_{i} - D_{i} K - 8\pi j_{i} = 0.\label{eq:constr_M}
\end{eqnarray}

A stable set of evolution equations (BSSN formulation~\cite{Shibata:1995we,Baumgarte:1998te}) can be obtained by the decompositions
\begin{eqnarray}
 \tgam_{ij} &=& \chi\gamma_{ij},\ \  \det\tgam_{ij}=1,\label{eq:def_tgam}\nonumber\\
 K_{ij} &=& \chi^{-1}\tA_{ij} + \frac{1}{3}\gamma_{ij}K,\label{eq:decompose_extcurvature}\nonumber\\
 \tGam^{i} &\equiv& -\partial_j\tgam^{ij},\label{eq:def_Gamma}
\end{eqnarray}
yielding
\begin{eqnarray}
 \ldif \chi
  &=&\frac{2}{3}\chi\left(\alpha K-\del_i\beta^i\right)\label{eq:evol_chi},\nonumber\\
 \ldif \tgam_{ij}
  &=&-2\alpha \tA_{ij}+\tgam_{il}\del_j\beta^l +\tgam_{jl}\del_i\beta^l
  -\frac{2}{3}\tgam_{ij}\del_l\beta^l\label{eq:evol_tgam},\nonumber\\
 \ldif K
  &=&\alpha\left[\tA_{ij}\tA^{ij}+\frac{1}{3}K^2
	    +4\pi\left(\rho+S\right)\right]-D^iD_i\alpha\label{eq:evol_K},\nonumber\\
 \ldif\tA_{ij}
  &=& \chi
  \Bigl[ \alpha\left( R_{ij} -8\pi S_{ij}\right) -D_iD_j\alpha \Bigr]^{TF}\nonumber\\
 & & +\alpha\left(K\tA_{ij}-2\tA_{il}\tA^l_j\right)
  \nonumber\\
 & & +\tA_{lj}\del_i\beta^l+\tA_{il}\del_j\beta^l
  -\frac{2}{3}\tA_{ij}\del_l\beta^l,\label{eq:evol_tA}\nonumber\\
 \ldif \tGam^{i\ } &=& 
  2\alpha\left(\tGam^i_{jk}\tA^{jk}
	      -\frac{3}{2}\frac{\del_j\chi}{\chi}\tA^{ij}
	      -\frac{2}{3}\tgam^{ij}\del_jK
	      -8\pi\tgam^{ij}j_j\right)\nonumber\\
 & &
  -2\tA^{ij}\del_j\alpha
  +\tgam^{jk}\del_j\del_k\beta^i
  +\frac{1}{3}\tgam^{ij}\del_j\del_k\beta^k\nonumber\\
 & &-\tGam^j\del_j\beta^i+\frac{2}{3}\tGam^i\del_j\beta^j
  -16\pi\alpha\tgam^{ij}j_j,
 \label{eq:evol_tGam}
\end{eqnarray}
where the superscript TF denotes the trace-free part, e.g.
$\displaystyle [R_{ij}]^{TF}\equiv R_{ij}-\frac{1}{3}\gamma_{ij}R$.\\

In addition, when we focus on a spherically symmetric geometry for single oscillatons,
we take the following metric ansatz instead of Eq.~\eqref{eq:line_element},
\begin{subequations}
 \label{eq:Ansatz_SS}
  \begin{align}
   \label{eq:Ansatz_SS_met}
   \dif s^2 =& -\alpha^2\dif t^2 +\psi^4\eta_{ij}\dif x^i\dif x^j\,,\\
   \label{eq:Ansatz_SS_ext}
   K_{ij} =& \frac{1}{3}\psi^4\eta_{ij}K\,,
  \end{align}
\end{subequations}
where $\eta_{ij}$ is the Minkowski 3-metric in spherical coordinates and $K_{ij}$ is the extrinsic curvature of the conformally flat metric $\gamma_{ij}=\psi^4\eta_{ij}$.
The constraints~\eqref{eq:constr_M} become
\begin{align}
  \psi^{-5}\Lap\hspace{-2pt}\psi 
 -\frac{K^2}{12}
 +\pi \left[\Pi^2 +\psi^{-4}\phi'^2+\mu^2\phi^2+\frac{1}{2}\lambda\phi^4\right] &=&\; 0\,,\nonumber\\
 \label{eq:Constr_SS}
 \frac{2}{3}K' +8\pi\Pi\phi' &=&\; 0\,,
\end{align}
and the evolution equations are explicitly described by
\begin{align}
 \dot\psi =& -\frac{1}{6}\alpha\psi K\,,\nonumber\\
 \dot K =& -\frac{\Lap\hspace{-2pt}\alpha}{\psi^{4}}
 -\frac{2}{\psi^{5}}\psi'\alpha'
 +\frac{\alpha}{3} K^2
 +4\pi\alpha\left(2\Pi^2 -\mu^2\phi^2-\frac{\lambda}{2}\phi^4\right)\,,\nonumber\\
 \dot\phi =& -\alpha\Pi\,,\nonumber\\
 \dot\Pi =& \alpha\Pi K -\frac{\alpha}{\psi^{4}}\Lap\hspace{-2pt}\phi
 -\frac{1}{\psi^{4}}\alpha'\phi'
 -\frac{2\alpha}{\psi^{5}}\psi'\phi'
 \label{eq:evol_SS}
 +\alpha\mu^2\phi+\alpha\lambda\phi^3\,,
\end{align}
where $\Pi\equiv-\dot{\phi}/\alpha$ is the momentum conjugate of $\phi$, a dot and a prime
denote derivative with respect to $t$ and $r$, respectively, and $\Lap$ is the flat Laplacian operator. 
The system must be closed by prescribing a gauge condition for the lapse function. We choose the $1+\log$ slicing condition defined by
\beq
\dot{\alpha}=-2\alpha K\,.
\eeq
To ensure that the boundary conditions are not contaminating the results, the outer boundary of the numerical domain is placed
sufficiently far away.

\subsection{Initial data}
\textbf{Single oscillaton:}
Initial data for a single oscillaton can be constructed by choosing $\phi=0$, the maximal slicing condition $K=0$ and considering the ansatz~\cite{Okawa:2013jba}
\be
 \Pi = \frac{A}{2\pi}\psi^{-\frac{5}{2}}\exp\left\{-\frac{r^2}{w^2}\right\},
 \quad
 \psi = 1 +\frac{u(r)}{\sqrt{4\pi}r}\,,\label{eq:ID_single}
\ee
where $A$ and $w$ are constants denoting the amplitude and width of the initial scalar pulse and we assume that the scalar field is initially centered around $r=0$.
The Hamiltonian constraint, Eq.~\eqref{eq:Constr_SS}, reduces to an ordinary differential equation
\be
 u''(r) + \frac{A^2r}{\sqrt{4\pi}}\exp\left\{-\frac{2r^2}{w^2}\right\}=0\,.
\ee
A particular solution, which is regular at infinity, is given by
\be\label{data_single}
 u(r) =\frac{A^2w^3}{16\sqrt{2}}\erf{\left(\frac{\sqrt{2}r}{w}\right)}.
\ee
The ADM mass can then be computed using
\be
M_T=\frac{u(r\rightarrow\infty)}{\sqrt{\pi}}=\frac{A^2w^3}{16\sqrt{2\pi}}\,.
\ee

\textbf{Collision of two oscillatons:}
We choose again $\phi=0$, the maximal slicing condition $K=0$, and we take the ansatz
\beq
 \Pi &=& \Pi_1 + \Pi_2,\quad
 \Pi_i = \frac{A_i}{2\pi}\psi^{-\frac{5}{2}}\exp\left\{-\frac{r_i^2}{w^2_i}\right\},\nn\\
 \psi &=& 1 +\frac{u_1(r)}{\sqrt{4\pi}r_1} +\frac{u_2(r)}{\sqrt{4\pi}r_2}
 +\frac{u(r)}{\sqrt{4\pi}r}\,,
\eeq
where $A_i,\, r_i=\sqrt{x^2+y^2+(z-z_i)^2}$ and $w_i$ are constants denoting the amplitude, location and width of the initial scalar pulses.
With this ansatz, and setting $z_1=w_1^2 L$ and $z_2=-w_2^2 L$ to cancel linear terms in $z$,
the Hamiltonian constraint reduces to three ordinary differential equations
\begin{align}
  u_1''(r_1) + \frac{A_1^2r_1}{\sqrt{4\pi}}\exp\left\{-\frac{2r_1^2}{w_1^2}\right\}&=0\,,\nonumber\\
  u_2''(r_2) + \frac{A_2^2r_2}{\sqrt{4\pi}}\exp\left\{-\frac{2r_2^2}{w_2^2}\right\}&=0\,,\nonumber\\
  u''(r) + \frac{\bA^2r}{\sqrt{4\pi}}\exp\left\{-\frac{2r^2}{w^2}\right\}&=0\,,
\end{align}
where $2/w^{2}\equiv(w_1^{2}+w_2^{2})/(w_1^2w_2^2)$ and $\bA\equiv\sqrt{2A_1A_2\exp\left\{-(w_1^2+w_2^2)L^2\right\}}$.

Particular solutions, analogous to the single oscillaton initial data~\eqref{data_single}, can be found and are given by
\begin{align}
 u_1(r) =& \frac{A_1^2w_1^3}{16\sqrt{2}}\erf{\left(\frac{\sqrt{2}r_1}{w_1}\right)},\nonumber\\
 u_2(r) =& \frac{A_2^2w_2^3}{16\sqrt{2}}\erf{\left(\frac{\sqrt{2}r_2}{w_2}\right)},\nonumber\\
 u(r) =& \frac{\bA^2w^3}{16\sqrt{2}}\erf{\left(\frac{\sqrt{2}r}{w}\right)}.
\end{align}
Equal-mass oscillatons can then be obtained by setting $A_1=A_2=A$ and $w_1=w_2=w$.

\subsection{Single oscillaton}\label{ssec:single}
\begin{figure}[ht]
 \psfig{file=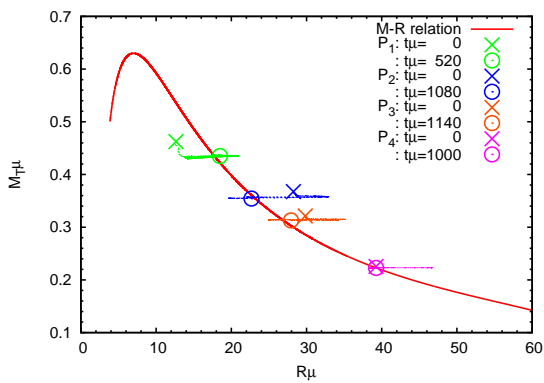,width=7.5cm}
 \caption[]{Path in the M-R relation curve, during time evolution of a single oscillaton using the Gaussican initial data as given by Eq.~\eqref{eq:ID_single}.
 Each color shows a different choice of parameters (amplitude and scalar mass given in Table~\ref{tab:param_init}).
 }
 \label{fig:Single_Oscillaton}
\end{figure}

%
In Fig.~\ref{fig:Single_Oscillaton}, we show that our initial data for single oscillatons, within an appropriate choice of parameters, result in single oscillatons.
We trace the time evolution for each initial data summarized in Table~\ref{tab:param_init}.
Note that collapse to a BH occurs when the amplitude of the scalar field is much
larger than the critical value discussed in Ref.~\cite{Okawa:2013jba},
while for very small amplitudes the field simply decays away to spatial infinity .
\begin{table}[h]
 \begin{tabular}{c|c}
  $Aw$ & $w\mu$ \\\hline
  $1.5$ & $8$ \\
  $0.9$ & $18$ \\
  $0.82$ & $19$ \\
  $0.6$ & $25$ \\
 \end{tabular}
 \caption{Initial data for a single oscillaton}
 \label{tab:param_init}
\end{table}

\subsection{Collision of two oscillatons}\label{ssec:collision}
\begin{figure}[ht]
 \psfig{file=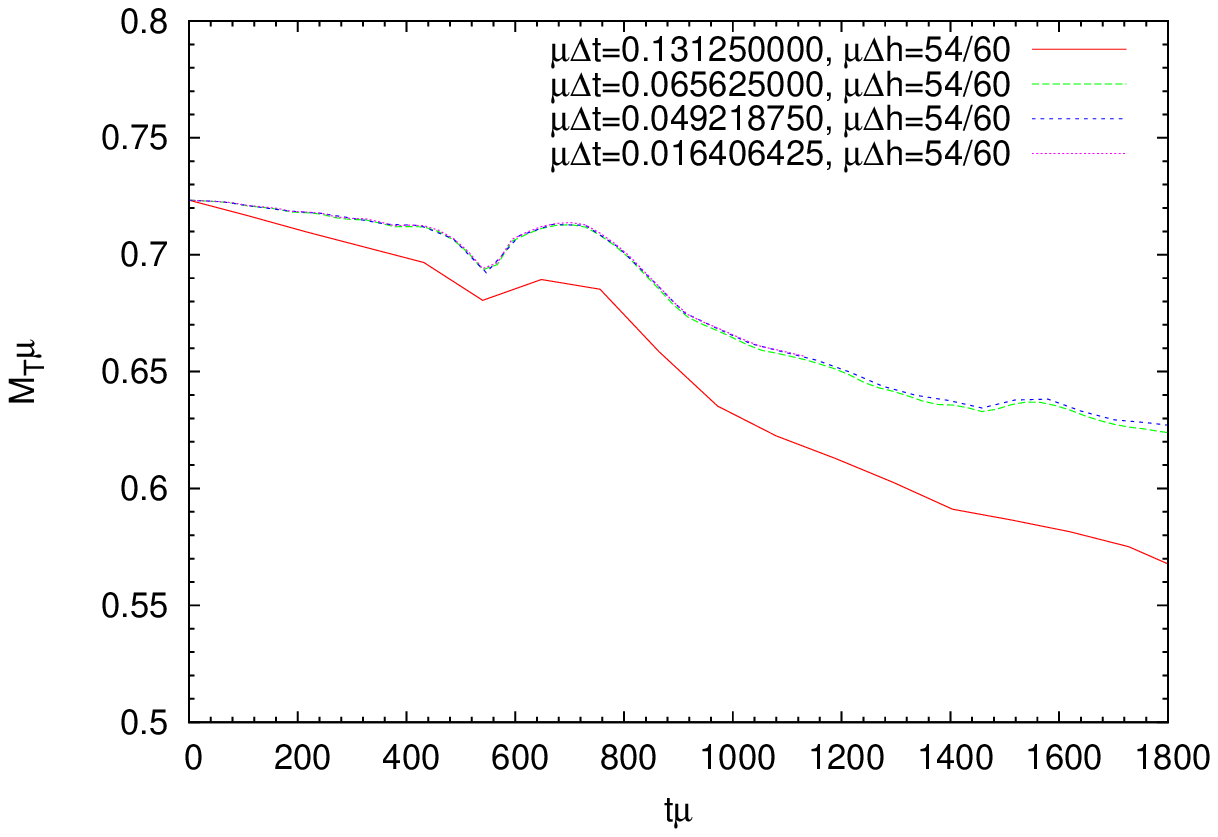,width=7.5cm}
 \psfig{file=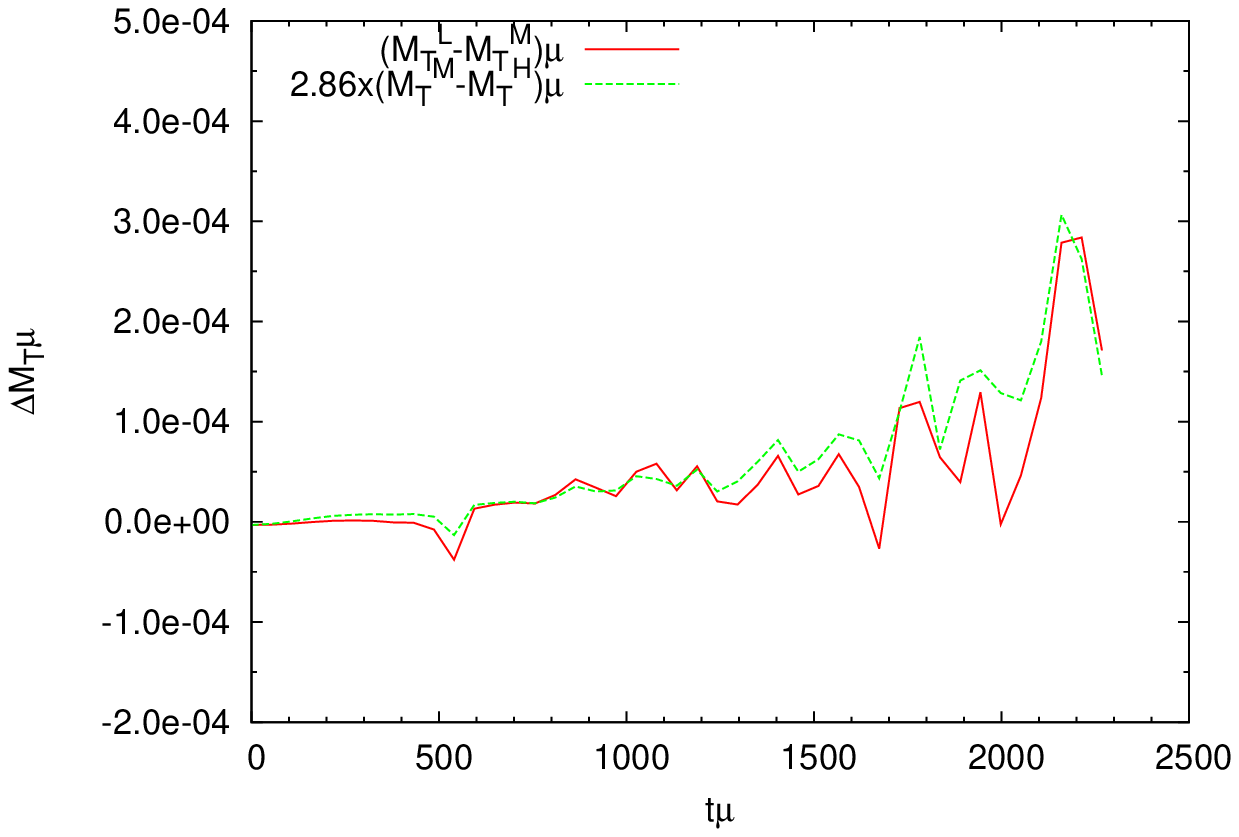,width=7.5cm}
 \caption[]{Time variation of total mass for the equal-mass collision with different resolutions.
 Initial data corresponds to the blue curve~$P_2$($Aw=0.9$ and $w\mu=18$) in Fig.~\ref{fig:Single_Oscillaton}.
 Upper: time resolution dependence. Two oscillatons merge at the time $t\mu\sim 600$.
 Lower: total mass difference among different spatial resolutions with $\mu\Delta t=0.065625$ fixed.
 The resolutions are set by $\mu\Delta h=54/40$ for lowest,
 $\mu\Delta h=54/60$ for middle and  $\mu\Delta h=54/80$ for highest resolution, respectively.
 The convergence factor is $\sim 2.86$ and then, the error is at most $0.1\%$ in our computational time.
 }
 \label{fig:Collision}
\end{figure}

Finally, we show the resolution study for the equal-mass collision in Fig.~\ref{fig:Collision}.
The upper figure shows the time resolution dependence
and it indicates that high-resolutions for the time scale of the
oscillation, $\mu\Delta t$, are needed in order to avoid a large dissipation due to insufficient resolution.
The lower figure shows the difference among total masses with different spatial resolutions as a function of time.
This shows that our error for the total mass is at most $0.1\%$ in our simulation.
The results in the main text are obtained using appropriate spatial and time resolutions.

\vskip 5cm

\bibliographystyle{h-physrev4}
\bibliography{Ref}

\end{document}